\documentstyle[12pt,aaspp4]{article}

\begin{document}
\def\Lya{Ly$\alpha$ }
\def\lya{Ly$\alpha$ }
\def\kms{km~s$^{-1}$ }
\def\cm2{\, \rm cm^{-2}}
\def\W#1{{W({\rm #1})}}
\def\N#1{{N({\rm #1})}}
\def\rAA{{\rm \, \AA}}
\def\sci#1{{\rm \; \times \; 10^{#1}}}
\def\Nav{N_a (v)}

\include{psfig}

\title{A KECK HIRES INVESTIGATION OF THE METAL ABUNDANCES AND
KINEMATICS OF THREE DAMPED {\Lya} SYSTEMS TOWARD
Q2206$-$199}

\author{JASON X. PROCHASKA \& ARTHUR M. WOLFE\altaffilmark{1}}
\affil{Department of Physics, and Center for Astrophysics and Space
Sciences; \\
University of California, San
Diego; \\
C--0111; La Jolla; CA 92093-0111\\
}

\altaffiltext{1}{Visiting Astronomer, W.M. Keck Telescope.
The Keck Observatory is a joint facility of the University 
of California and the California Institute of Technology.} 

\begin{abstract} 

	We present high resolution, high 
SNR spectra of the QSO Q2206-199 obtained with HIRES on the
10m W.M. Keck Telescope.  Our analysis focuses on the two
previously identified damped \lya systems found at $z=1.920$
and $z=2.076$.  For the $z=1.920$ system, we measure
accurate abundances (relative to solar) for Fe, Cr
Si, Ni, Ti, and Zn:
[Fe/H] = $-0.705 \pm 0.097$, [Cr/H] = $-0.580 \pm 0.100$,
[Si/H] = $-0.402 \pm 0.098$, [Ni/H] = $-1.012 \pm 0.095$,
[Ti/H] = $-0.776 \pm 0.081$,  and
[Zn/H] = $-0.379 \pm 0.097$. 
This system exhibits the highest metallicity we have measured for
a damped \lya system.
By contrast the $z=2.076$ system is
the most metal poor ([Zn/H] $< -1.745$) we have analyzed,
showing absorption features
for only the strongest transitions.  We determine accurate
abundances for Fe, Si and Al:
[Fe/H] $= -2.621 \pm 0.071$, [Si/H] $= -2.225 \pm 0.075$,
and [Al/H] $= -2.727 \pm 0.070$.

	Analyses of the abundance variations 
of Fe, Ni, Cr, and Si relative to Zn and the abundance
trends versus condensation temperature do not offer positive 
evidence for the presence of dust in the damped system at $z=1.920$.  
In addition, the relative abundance ratio [Ti/Fe] $\approx 0$
further suggests the absence of ISM-like dust.
Unfortunately, the lack of measurable Zn absorption
in the $z=2.076$ system does not allow a similar investigation for 
the presence of dust.

In addition to the significant
difference in metallicity, the two damped systems have vastly different
kinematic characteristics.  The $z=1.920$ system 
spans $\approx 150$ \kms
in velocity space (measured from the low-ion transitions)
while the $z=2.076$ system spans a mere 30 \kms.  Furthermore, the
$z=1.920$ profiles are significantly asymmetric while the $z=2.076$
profiles are highly symmetric.  Even with these differences, we
contend the two systems are consistent with one physical description,
that of a thick, rotating disk. 

	In addition to the two previously identified damped \lya systems,
we investigate a very strong Mg II system at $z=0.752$ which is 
very likely yet a third damped \lya system along the line of sight.
The very weak Mn II 2606,
2594, 2576 and Ti II 3073, 3342, and 3384 transitions have been
positively measured and suggest a very conservative lower limit H I
abundance of $\log \N{HI} > 19.0$ assuming metallicity relative to
solar equal to 0 and no depletion.  Together with the damped system at
$z = 1.920$, this marks the first confident ($> 5 \sigma$) detection
of Ti in QSO absorption line systems.  We analyze the abundance
ratios [Mn/Fe] and [Ti/Fe] and their values are inconsistent with
dust depletion, yet consistent with the abundance pattern detected
for halo stars in the Galaxy (\cite{lu96a}).

	Finally, we identify a C IV system at $z=2.014$ that shows
a very narrow feature in Si IV and C IV absorption.  The corresponding $b$
values (5.5 \kms and 8.9 \kms for Si IV and C IV respectively) for
this component
suggest a temperature of $4.7 \sci{4} \; \rm K$.  Because collisional 
ionization can explain the observed abundances only
for $T > 8 \sci{4} \; \rm K$, we contend
these ions must have formed through a different physical
process (e.g. photoionization).

\end{abstract}


\section{INTRODUCTION}

  This paper is the third  in a series devoted to studying 
the metal content of
high-redshift galaxies and their progenitors with the 10m W.M. Keck 
Telescope.  Our primary objectives are

\indent (1) to record the emergence of metals in galaxies, 

\indent (2) to trace the mean cosmic metallicity from
$z$ $\approx$ 4.5 to the present,

\indent (3) to determine the kinematic state of galaxies
from $z$ $\approx$ 4.5 to the present. 

\noindent We are implementing this study using HIRES, the echelle
spectrograph on the Keck 10m Telescope (\cite{vgt92}), to obtain 
high-resolution spectra of QSOs with foreground damped {\Lya}
systems. The damped {\Lya} systems are a population of neutral
gas layers exhibiting properties indicating they 
are either galaxy progenitors, 
or well formed galaxies detected during an early evolutionary 
phase. Recent
studies indicate the comoving density of neutral gas in damped 
systems at $z
\approx 3.3$ is comparable to the density of 
visible stars in current galaxies.  At
lower redshifts, the comoving density of neutral gas 
decreases with time in a
manner consistent with gas consumption by star formation 
(Wolfe et al. 1995),
and at higher redshifts there is tentative evidence for an increase in 
comoving density with time
possibly indicating accretion of gas (\cite{lom95}).
Therefore, studies of the metal content of the damped
\Lya systems enable one to trace the 
chemical evolution of representative galaxies from a
presumably metal-poor gaseous progenitor phase to 
metal-rich epochs when most
of the baryons are in stars. As a result the 
age-metallicity relation, kinematic
conditions, etc., deduced from the damped \Lya systems should tell us more
about the history of galaxies at large 
redshifts than analogous relations deduced
from old stars found in the solar neighborhood (\cite{evd93}).

In previous papers we have examined two damped \Lya systems
at $z=2.309$ toward PHL 957 (\cite{wol94}) and 
at $z = 2.462$ toward Q0201+365 (\cite{pro96a}).  We derived
metallicities\footnote{Note that the use of Zn as
the indicator of metallicity is now in question.  See $\S \ref{metll}$
below.} from the Zn abundances of [Z/H] = $-1.55 \pm 0.11$
and $\rm [Z/H] > -0.562$ respectively.  Investigating the kinematics
of the two systems from their low-ion velocity profiles,
we noted a systematic edge-leading asymmetry where the majority
of gas is found in at one edge with decreasing absorption
toward the other edge.  We contend that this edge-leading
asymmetry is consistent with a thick rotating disk with rotation
speeds $v \approx 200$ \kms which resembles the kinematics
of our own Galaxy.  Finally, we estimated the dust content
of the damped system toward Q0201+365 by studying the variations
in the abundances of Fe, Si, Ni, and Cr relative to Zn (Zn
has very low levels of adsorption onto dust grains).  Our analysis
suggests this system is relatively free of dust grains.

     This paper presents HIRES spectra for Q2206$-$199, a $V$
= 17.3 QSO with emission redshift $z_{em}$ = 2.56.  We focus this
study on two damped \Lya systems at $z$ = 1.920 and $z$ = 2.076
examined previously by several authors (\cite{ber91,rau90,rob83,rau93}).  
Pettini et al. (1994) have fitted a Voigt 
damping  profile to the \Lya absorption troughs at $z = 1.920$ 
and $z = 2.076$ and found 
$\N{HI} = 4.5 \pm 0.8 \sci{20} \cm2$ and $\N{HI} = 2.7 \pm 0.4
\sci{20} \cm2$ respectively.  Because the system at $z = 1.920$ is 
very metal rich, relatively good measurements of the abundances
of Cr, Zn, and Si have been made with lower resolution 
data (\cite{ber91}, \cite{ptt94}).  That work has
a limited wavelength coverage and a relatively low signal to noise ratio
(SNR).  Therefore an analysis of the abundances of Ni and
Fe could not be successfully performed. 
Rauch et al. (1990) have examined the $z = 2.076$ damped \Lya
system with moderate resolution ($\approx 30$ \kms) and SNR.  Their
observations were limited to the \Lya forest, however, and therefore
suffered from contamination by \Lya clouds.
They did, however, confirm the very low abundance of
the $z = 2.076$ system first noted by Robertson et al. (1983).
Finally, Pettini et al. (1994) have placed upper limits on the
Zn and Cr abundances.  

	In the 
present study we obtain spectra at a resolution of  $\approx 8$ {\kms}
with a typical signal-to-noise ratio of 45:1 spanning $\approx 2500 \rAA$
from $\approx 4000 - 6500 \rAA$.  The high quality of the
spectra allows us to focus on weak unsaturated transitions 
of ions expected to dominate the ionization state
of gas in neutral clouds.  In addition, we are able to concentrate
on transitions outside of the \Lya forest.
In the following analysis of the $z = 1.920$
system, we establish accurate 
abundances relative to H for Ti, Fe, Ni, Si, Cr, and Zn and estimate the
abundances of Cu and Ge.  We also examine the abundance
of Fe, Si, Ni and Cr relative to Zn in order to investigate
the level of dust grain depletion.  For the metal poor damped \Lya 
system at $z = 2.076$, we accurately measure abundances for Fe, Si, and Al
and place a strict upper limit on the abundance of Zn and Ni.
In addition to the detailed abundance 
measurements, we investigate the kinematic characteristics associated
with the two damped \Lya systems.  Finally, we analyze
several significant Mg II and C IV systems identified toward Q2206-199.

\begin{figure}
\centerline{
\psfig{figure=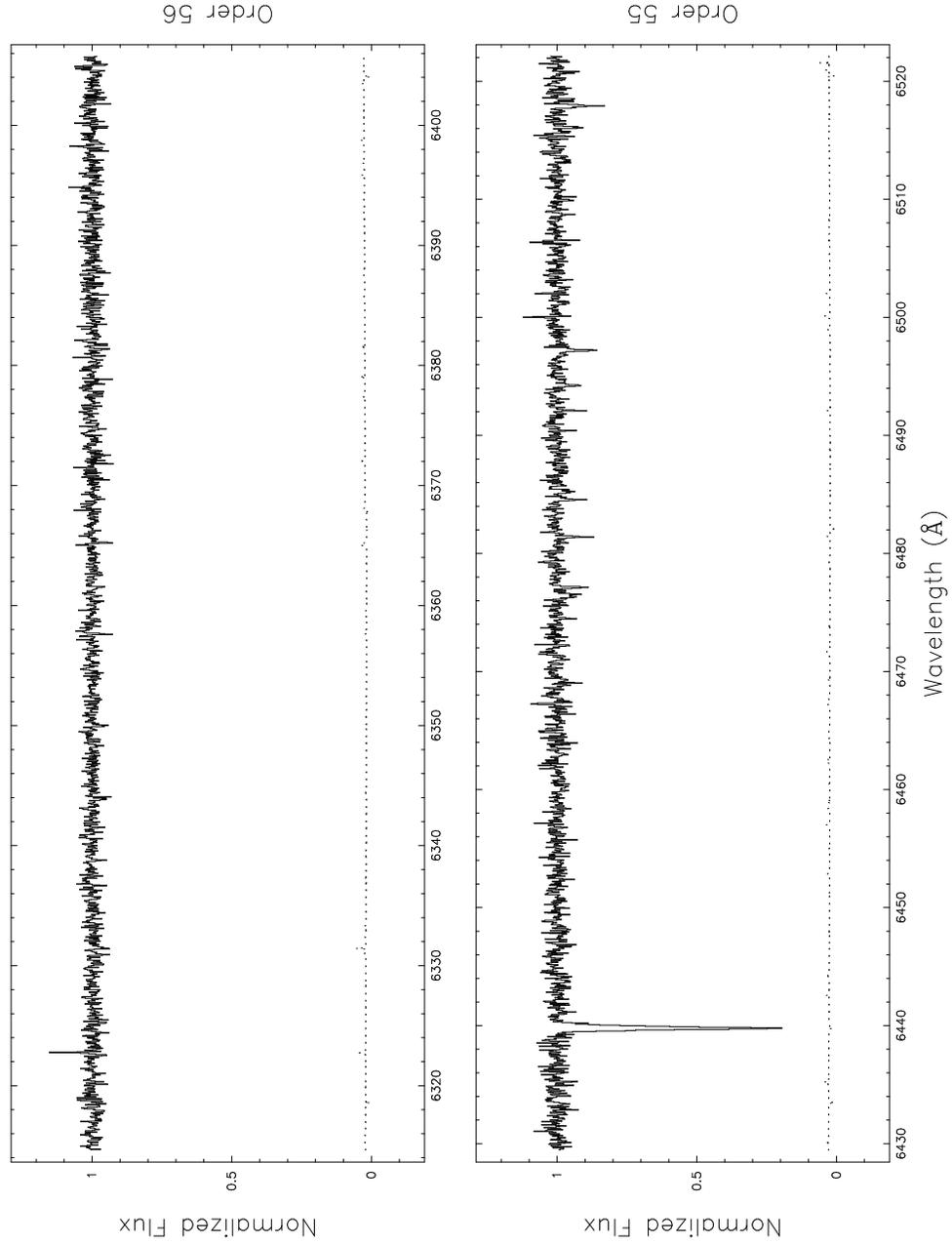,height=7.5in}}
\caption{ Keck HIRES spectra of Q2206-199 at a 
resolution of $\approx 8$ \kms
and a SNR of $\approx 45$.  All 36 orders presented 
are identified by the
echelle order intrinsic to the Keck HIRES spectrograph.  
The dotted line is a
1$\sigma$ error array derived assuming Poisson statistics.}
\label{sptra}
\end{figure}

	The paper is organized as follows.  
$\S 2$ briefly reviews the acquisition
and reduction of the HIRES data.  We present the spectra in 
Figure~\ref{sptra} and an extensive absorption 
line list in Table~\ref{abs}.
In $\S 3$, we present velocity profile plots and least-squares fits 
of the more extensive metal line systems.  We utilize the results of the
least-squares fits and the apparent optical 
depth method to determine ionic
column densities of the most significant metal line systems in $\S 4$.
$\S 5$ gives the abundances (measured relative to solar) of the
2 confirmed damped \Lya systems and a third possible damped \Lya system.
We also look for variations of the relative
abundances with respect to velocity in the $z = 1.920$ system.  
In $\S 6$, we investigate the kinematic characteristics of the 2 damped
\Lya systems, comparing them with one another as 
well as with previous results
for other damped \Lya systems.  Finally, $\S 7$ gives a 
summary of the results.

\section{DATA}

     In this section we present the HIRES spectra, detailing the
techniques used for the acquisition and reduction of the data. 
Table~\ref{abs} gives an absorption line list with measured equivalent
widths and 1$\sigma$ errors and identifies over 85$\%$ of the features
redward of the \Lya forest.

\subsection{Acquisition}

     We observed Q2206-199 with the HIRES echelle spectrograph
on the 10m W. M. Keck Telescope for 6.2 hours of total integration time
over three nights of observation.  A journal of 
the observation dates, exposure
times, wavelength coverages and resolution of the data are presented in 
Table~\ref{jou}.  We used the C5 decker plate with
1.1$''$ slit, the kv380 filter to eliminate 2nd order blue wavelengths, 
and standard 2$\times 1$ binning on the 2048 x 2048 Tektronix CCD. 
As Table~\ref{jou} indicates, we used two different CCD alignments
covering two different wavelength regions for our observations.  
The setup
provided a resolution (FWHM) ranging from 7.2 $-$ 8.0 \kms and wavelength
coverage from $3940 - 6520 \rAA$ over 36 orders.  
There are gaps
in the wavelength coverage between successive
orders for orders redward of $5100 \rAA$
where the free spectral range of the echelle exceeds the format of the CCD.
We also took $\approx 700$ s exposures of the standard star BD+28411,
1s Th-Ar arc images, dark (bias) frames and flat fields for 
calibration.

\subsection{Reduction}

	We reduced the data with the techniques discussed in 
Prochaska and Wolfe (1996a).  In short, we utilized a software package
kindly provided by T. Barlow (1996) to convert the 2-D CCD images
to uncalibrated 1-D spectra.  We then continuum fit the data
with Legendre polynomials and wavelength calibrated to
proper vacuum heliocentric wavelengths with the
Th-Ar arc frames and the standard IRAF package {\it id}.  We also
derived a 1$\sigma$ error array from the data 
assuming Poisson statistics while 
ignoring the errors associated with continuum fitting
(i.e. $\sigma = \sqrt{N_{\rm obj} + N_{\rm sky}}$).  Table~\ref{snr}
lists the average SNR for several orders of the spectra.  There is a
significant decrease in the SNR for wavelengths blueward of $\approx
4140 \rAA$ both because of the shorter total integration time
and the insensitivity of HIRES at bluer wavelengths.  
Figure~\ref{sptra} presents the spectra (solid line) along with 
the 1$\sigma$ error array (dotted line).

\subsection{Absorption Lines}

     Table~\ref{abs} lists the wavelengths, equivalent widths and 
1$\sigma$ errors for all absorption line features which 
exceed the 5$\sigma$ limit in equivalent width as measured by 
techniques similar to those of Lanzetta et al. (1991).  
The reported wavelengths represent rough estimates of the centroids 
of complex line profiles and should be taken only as guides 
for differentiating between features.  We do not report
equivalent widths for those features identified as sky absorption lines or
complicated multiple transition blends. 

     Table~\ref{abs} includes the transition names and approximate
redshifts of those features we successfully identified.  
Identification proceeds in a
largely {\it ad hoc} fashion, with the initial emphasis placed on
finding C IV, Mg II and Si IV doublets.  
Having composed a list of redshifts for the
metal-line systems, we attempt to match the remaining features with the
strongest metal-line transitions.  
Finally, we compare the line profiles of the
individual transitions of the redshift systems in 
velocity space for conclusive identification.  Note that the vast 
majority of unidentified features blueward 
of $4350 \rAA$ are presumed to be \Lya lines.

     By comparing the object frames with the identically reduced
sky and standard star images, we identified night-sky 
absorption features in the spectra.  These features are labeled
appropriately in Table~\ref{abs} along with all other identified spurious
features.

\section{ANALYSIS}

	This section presents the velocity profiles for the most
complex metal line absorption systems toward Q2206-199.  
In all plots, a dot-dashed vertical line is drawn for reference,
usually identifying the strongest feature
present.  For clarity, we have plotted features not related 
to the given system
with dotted lines.  Table~\ref{osc} lists the adopted
wavelengths and oscillator strengths taken from Morton (1991) with the
exception of Si, Zn, Cr and Fe where we adopt oscillator
strengths and wavelengths
from Tripp et al. (1995) and Cardelli and Savage (1995).

For several
of the systems, we superimpose the solutions of the least-squares
fits from the VPFIT package kindly provided by R. Carswell.  The
VPFIT package is a least-squares program which minimizes the
$\chi^2$ matrix of a multiple component Gaussian fit to line profiles.
The package allows one to tie the redshift and b values (where
the Doppler velocity $b$ and velocity dispersion are 
related by $b = \sqrt{2} \sigma$) of a given component from ion to
ion while allowing the column densities to vary individually.
The package calculates errors in the fitted parameters based on
both the SNR of the data and the quality of the fit.

\subsection{Damped \Lya Systems}

\subsubsection{$z$ = 1.920}

\begin{figure}
\centerline{
\psfig{figure=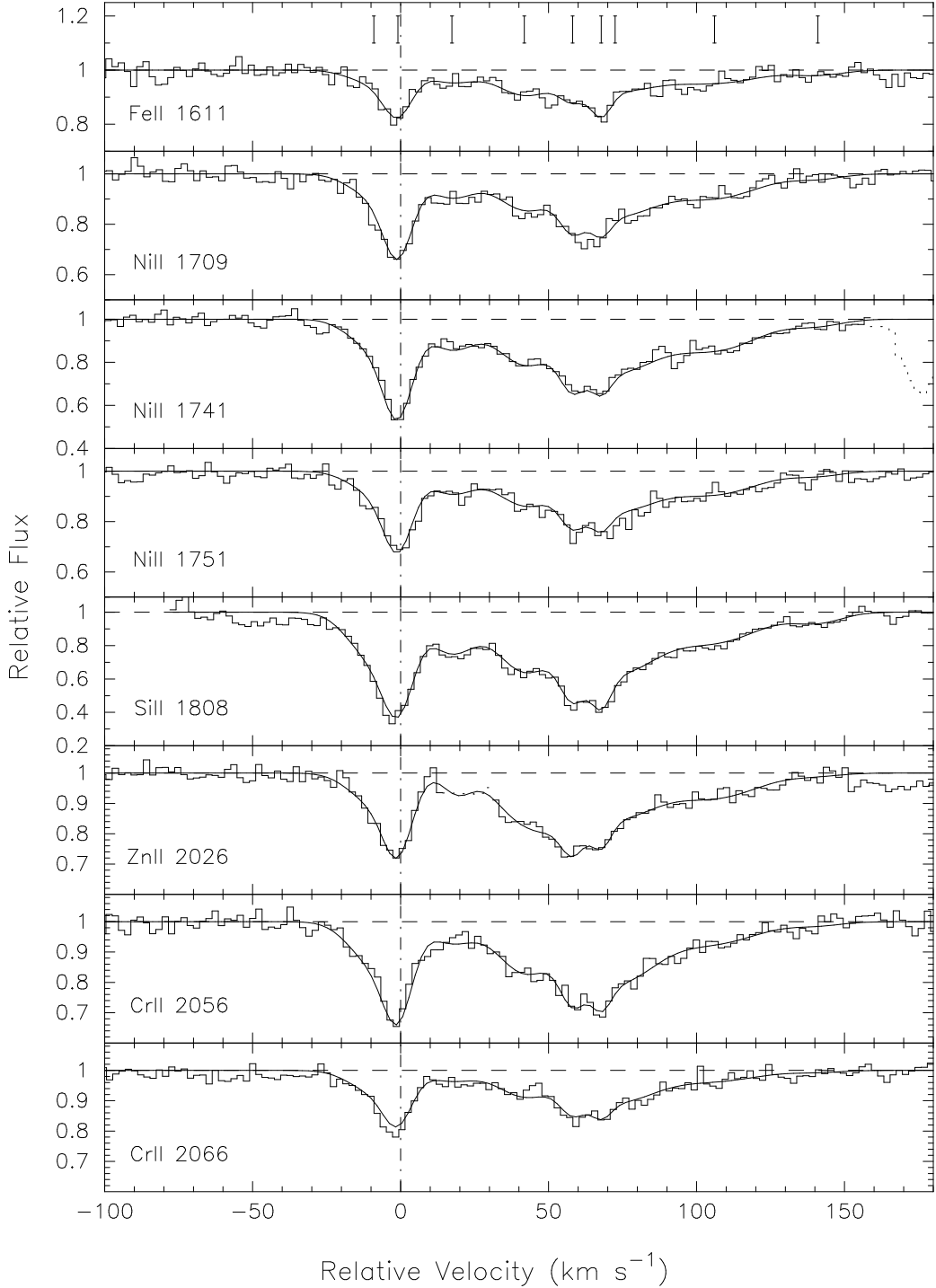,height=7.5in}}
\caption{ Velocity profiles of the low-ion transitions from 
the system at $z$=1.920.  
The overplotted curve represents the least-squares
VPFIT solution to the 8 transitions.
The dotted profile indicates blends from intervening systems.
The dashed vertical line corresponds to $z$=1.920.
The marks above the Fe II 1611 indicate the 
velocity centroids for the 
VPFIT solution as listed in Table 5a.
In all figures with VPFIT solutions, the leftmost mark is component 1.}
\label{1920V}
\end{figure}

     Figure~\ref{1920V} presents the velocity profiles and VPFIT
solutions of the low-ion transitions associated with
the damped \Lya system at $z$=1.920.  
The velocity centroids for the Gaussian
components of the fit are denoted by the short vertical lines
above the Fe II 1611 profile.  In our solution, we tied the 
$b$ values and redshifts of each component from ion to ion and allowed 
for individual variations only in the column densities.  
We found that a minimum of 9 individual
components were required to optimally fit the 
low-ion transitions.  Our solution yields a reduced 
$\chi_\nu^2 = 1.057$ with a corresponding probability $P_{\chi^2} = 0.152$.  
The majority of the error in the fit lies in the Ni II profiles, most
likely due to hidden saturation at $v \approx 65$ \kms 
(see $\S \ref{shca}$).

	We have removed a feature in the Zn II profile
at $v = 17$ \kms (indicated by the dotted line)
identified as a poorly subtracted sky line.  In addition, we have 
accounted for absorption from Mg I 2026 by introducing a single Mg I
component
(at $v = 0$ \kms in the Mg I 2026 frame) in the VPFIT solution.
Table~\ref{1920L} lists the redshift, $b$ value and column density 
along with 1$\sigma$ error of every velocity component 
in the VPFIT solution.  The quality
of the fit reflects the high degree to which the low-ion 
profiles track one another.

\begin{figure}
\centerline{
\psfig{figure=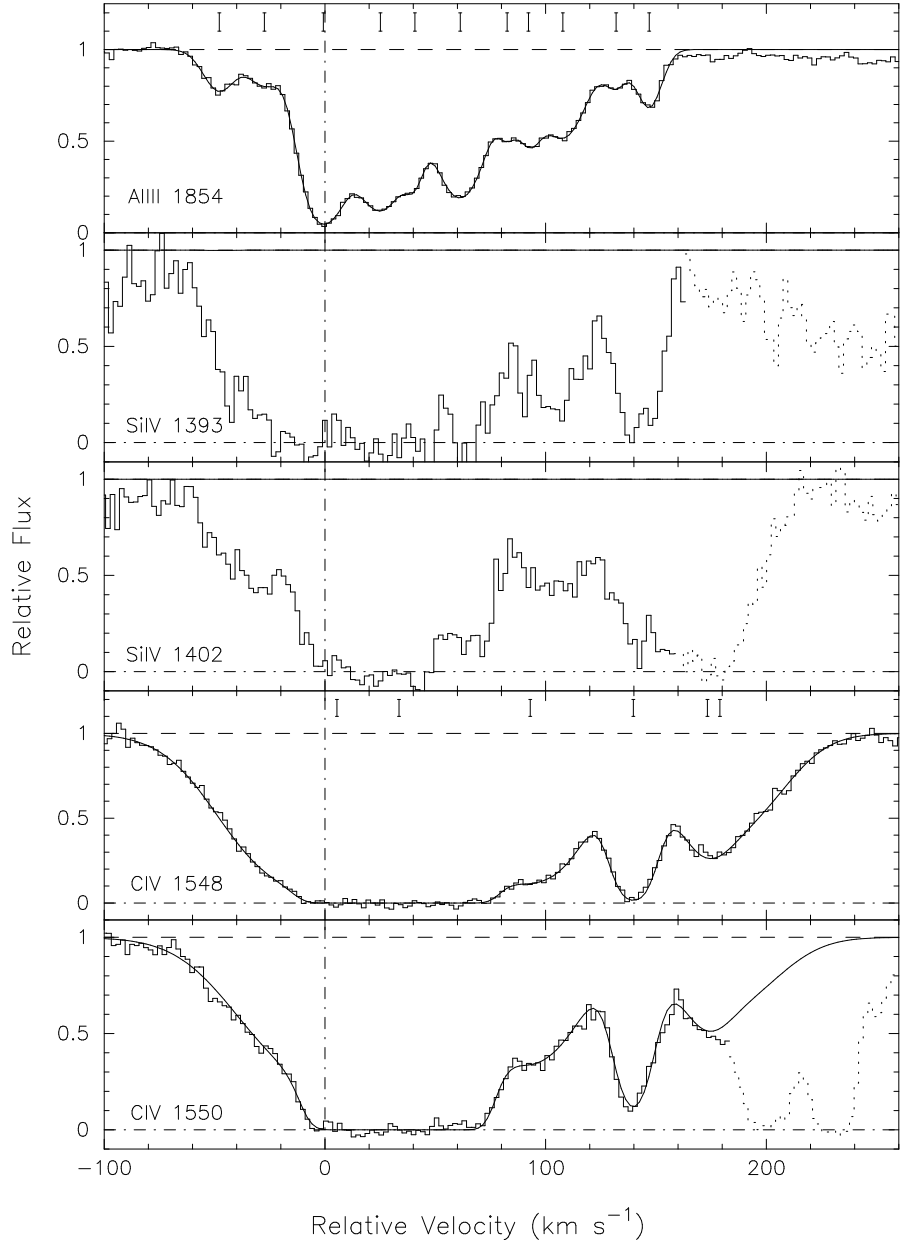,height=7.5in}}
\caption{ Velocity profiles and VPFIT solutions of the high-ion and Al III
transitions from the system at $z$=1.920.
The dashed vertical line corresponds to $z$=1.920.  The dotted
profile indicates blends from intervening systems.}
\label{1920B}
\end{figure}

	The VPFIT solutions and velocity profiles for the high-ion
and Al III transitions are given in Figure~\ref{1920B}.  In this case
the vertical lines above the Al III 1854 and C IV 1548 indicate
the velocity centroids of the Gaussian components for the Al III and 
C IV solutions respectively.
The Al III profile does track the low-ion profiles (much more so
than the high-ion profiles) but not closely
enough to yield a successful simultaneous solution.   In particular, 
absorption observed in the Al III profile at $v \approx 30$ and 140
\kms is not evident in the low-ion profiles.  
The results of the fits for Al III (and C IV) 
are listed in Table~\ref{1920H}.
The Al III solution
yields a reduced $\chi_\nu^2 = 1.00$ with a probability
$P_{\chi^2} = 0.5$ as expected for a one ion solution.

Although the high and
low-ion transitions span the same velocity interval, the profiles
are significantly different.  
 For instance, the C IV profile is
much less resolved and shows less absorption
at $v \approx 100$ \kms.  We chose not to fit Si IV because of its very
poor SNR and blends with several \Lya lines. 
Note that the features at $v \approx 200$ \kms in the C IV 1550
profile are blends with Fe II 2586 from the $z = 0.752$ system and
were included in the C IV fit.  The solution yields a relatively
high reduced $\chi_\nu^2 = 1.64$ which is entirely due to the
blending in the C IV 1550 profile.  In contrast, the reduced
$\chi_\nu^2$ value for the C IV 1548 profile alone is less than 1.

\begin{figure}
\centerline{
\psfig{figure=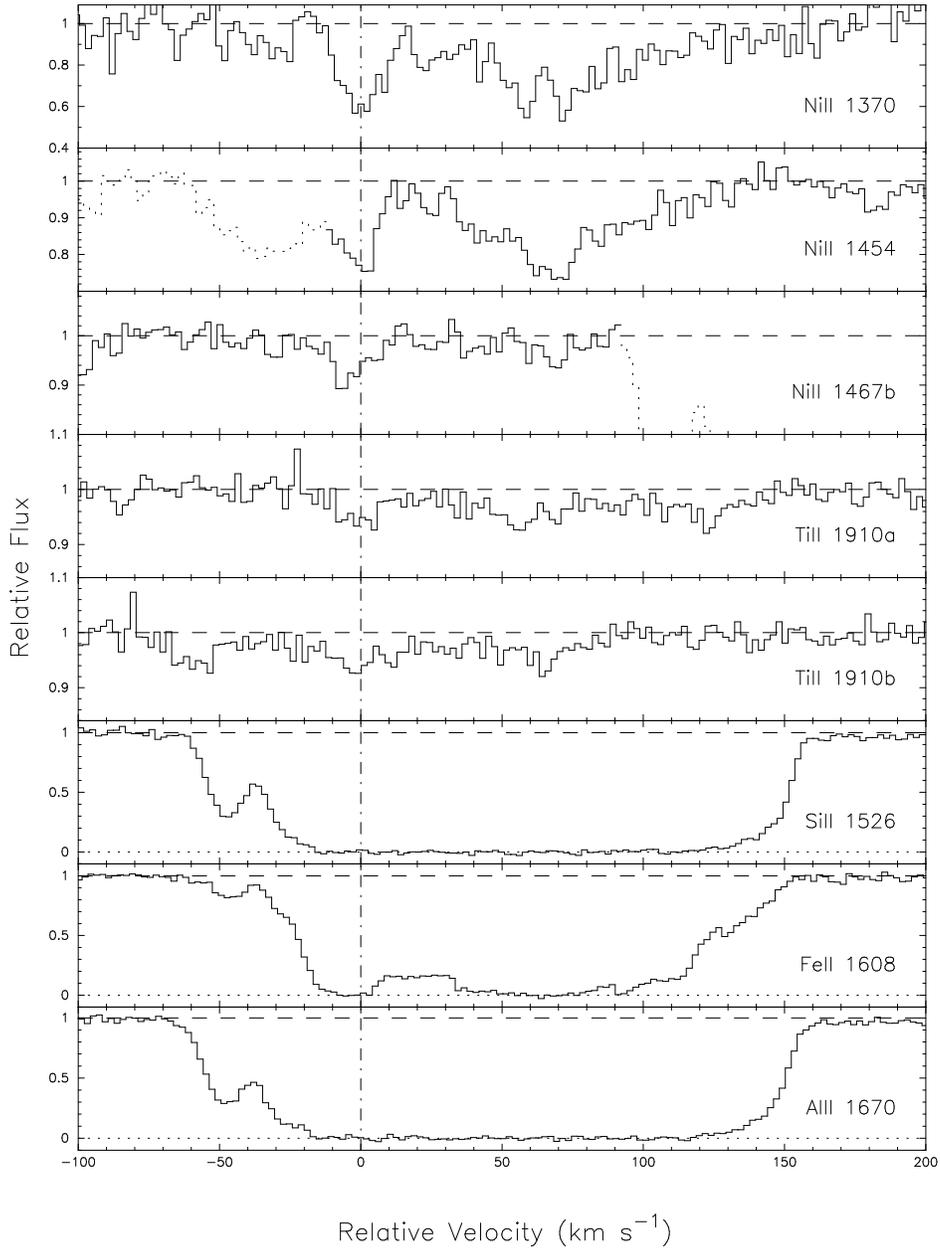,height=7.5in}}
\caption{ Velocity profiles of those metal-line transitions from the
system at $z=1.920$ without VPFIT solutions.  The dashed vertical
line is at $z$=1.920.}
\label{1920N}
\end{figure}

	Those velocity profiles not fit with the VPFIT package
are presented in Figure~\ref{1920N}.  
The three Ni II transitions are found in the \Lya forest 
and therefore suffer from blending and poor SNR.  The Ti II 1910
transitions are blended with one another and are very weak. 
The Si II 1526, Fe II 1608 and Al II 1670 profiles
are all saturated and therefore cannot be fitted successfully.

\subsubsection{$z$=2.076}

\begin{figure}
\centerline{
\psfig{figure=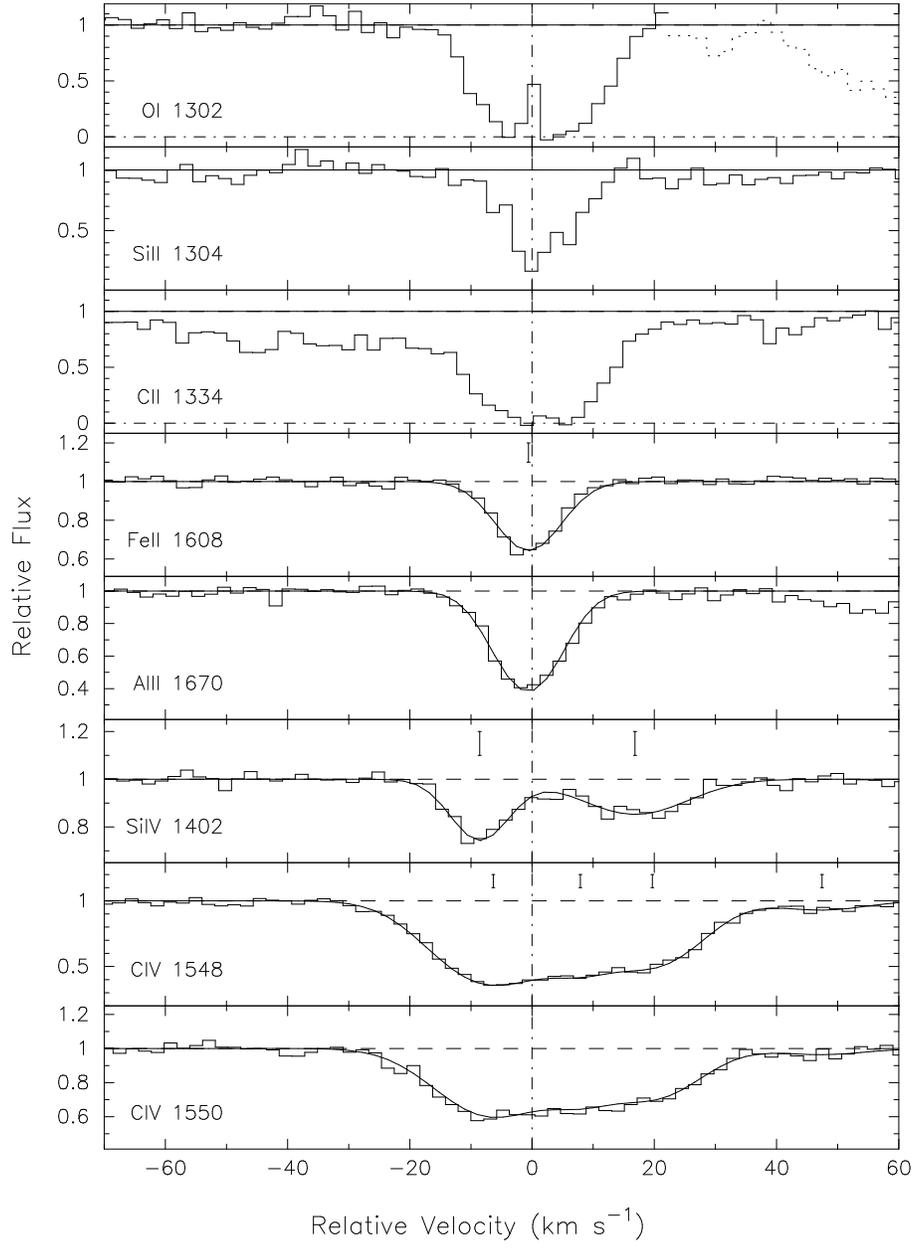,height=7.5in}}
\caption{ Velocity profiles and VPFIT solutions for the 
transitions from the system
$z$=2.076.  The dashed vertical line corresponds to $z$=2.07623. }
\label{2076V}
\end{figure}

	Figure~\ref{2076V} shows the velocity profiles and VPFIT
solutions (where applicable) of the metal line transitions associated
with the damped \Lya system at $z = 2.076$.  As discussed below,
this system has a very low metallicity and therefore exhibits very
few metal line transitions. 
In the VPFIT solution for the low ions Fe II 1608 
and Al II 1670 (with velocity centroids denoted above the
Fe II 1608 profile) we found a 1 component fit was optimal.  Although
two or more components are likely present in the feature,
the resolution is insufficient to resolve them.  
Introducing a second component overdetermines the
solution and leads to large errors.  We chose not to fit
the other low-ion profiles because they lie in the \Lya forest
and therefore have low SNR and possible \Lya contamination.  
Table~\ref{2076F} presents the results of the low-ion solution.

	In contrast to the low-ions, the high-ion solution for C IV
required 4 components and the Si IV solution 2 components (denoted by
the vertical lines above the C IV 1548 and Si IV 1402 profiles). 
The $b$ values and redshifts differ significantly between the two 
solutions as well as with the low-ion solution.  
Thus there appears to be
an inherent difference between the high and low-ions as well as 
between the individual high-ions.  This may be a result of absorption
from different regions within the system or the presence of a
multi-phase medium.  The fits were both successful,
$\chi^2_\nu = 0.654$, $P_{\chi^2} = 0.977$ for Si IV and 
$\chi^2_\nu = 1.032$, $P_{\chi^2} = 0.401$ for C IV.
The results are given in Table~\ref{2076F}.

\subsection{Mg II Systems}

\subsubsection{$z$ = 0.752}

\begin{figure}
\centerline{
\psfig{figure=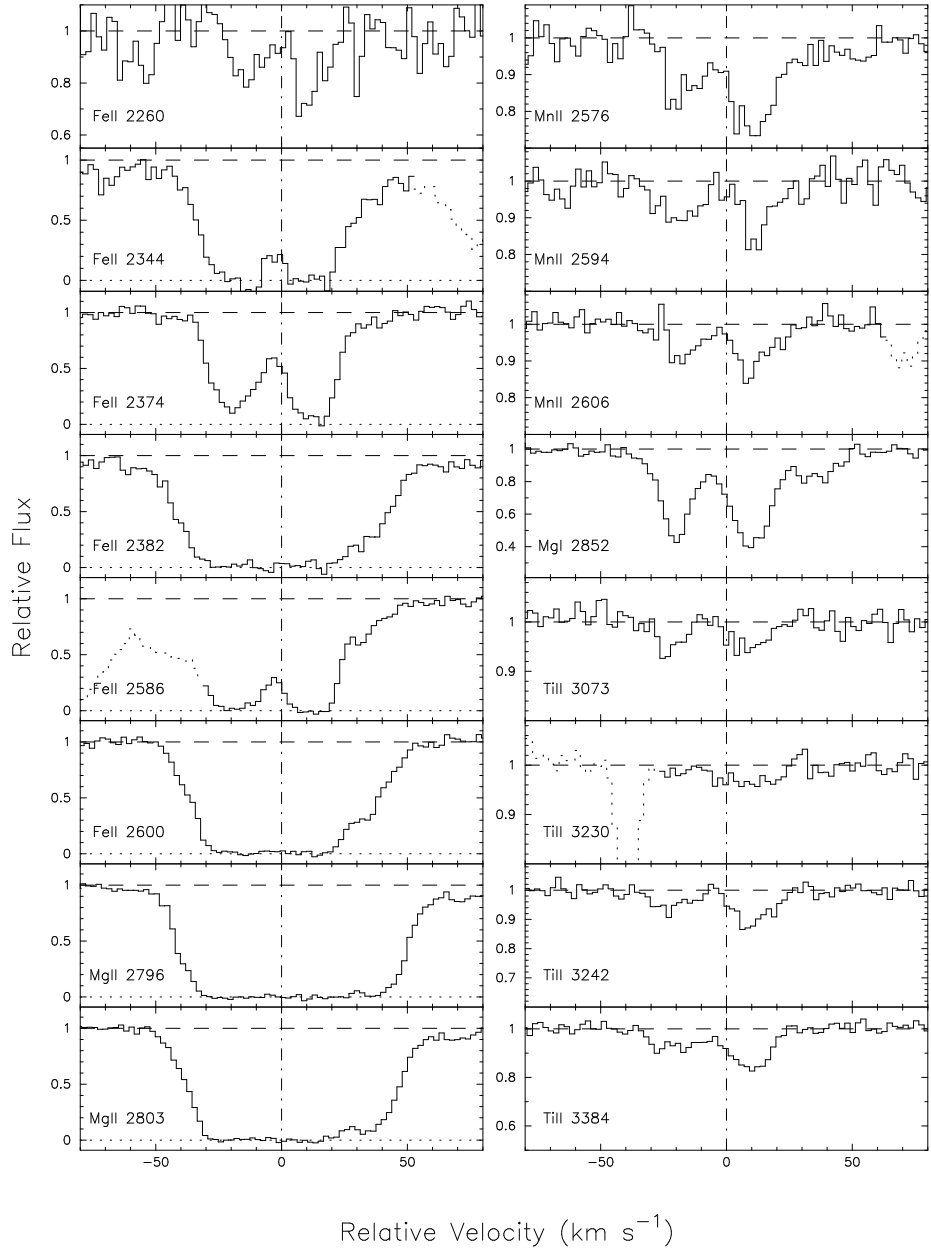,height=7.5in}}
\caption{ Velocity profiles of the Mg II metal system at 
$z$=0.752.  The dashed vertical line is at $z$=0.7519.}
\label{0752V}
\end{figure}

	Figure~\ref{0752V} shows the velocity plots 
of the very strong Mg II system at $z$
= 0.752.  The system is so metal rich that nearly all of 
the Fe II and Mg II transitions are saturated.  Conversely, the Mn II 
and Ti II transitions are so weak they are dominated by
noise.  Therefore we chose not to perform a least-squares fit.

\subsubsection{$z $ = 0.948}

\begin{figure}
\centerline{
\psfig{figure=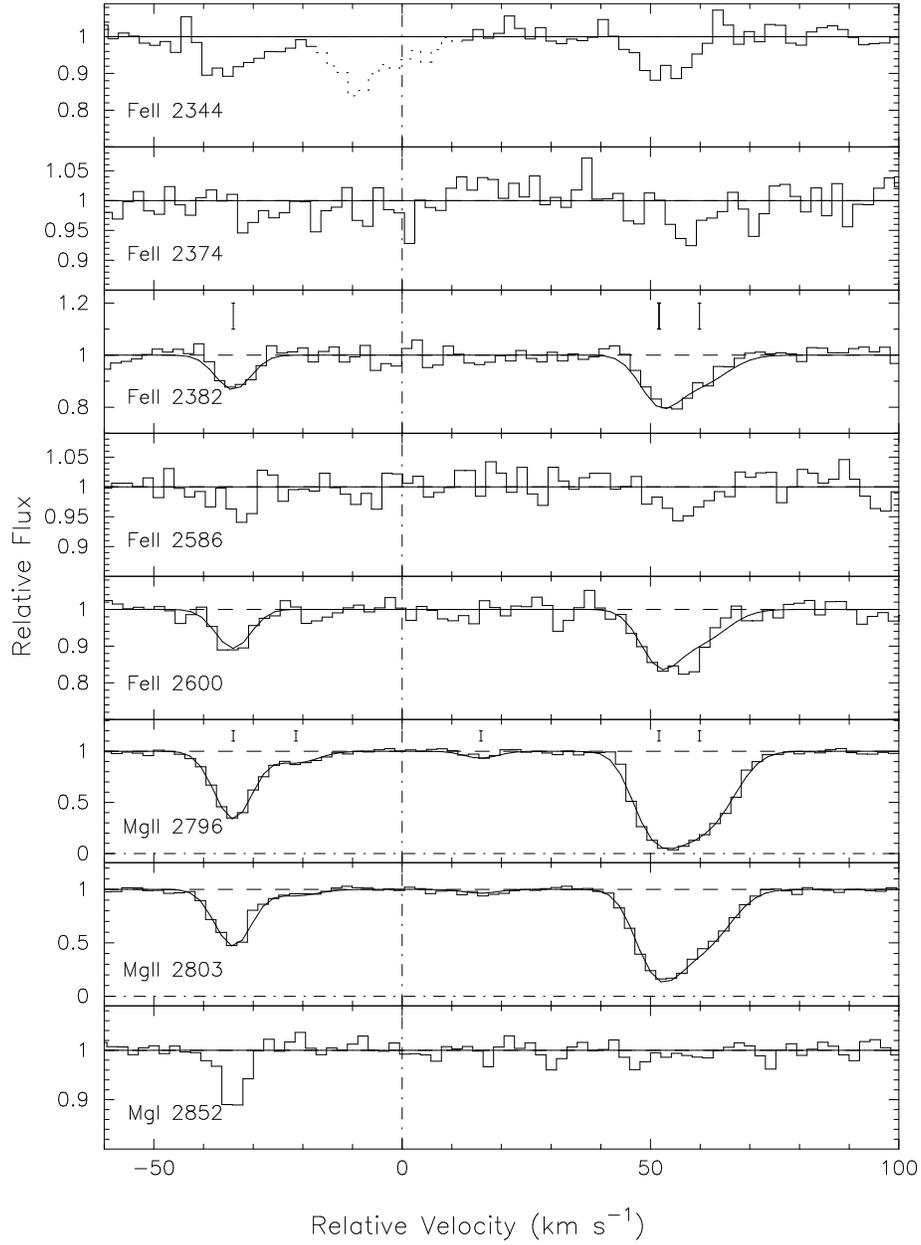,height=7.5in}}
\caption{ Velocity profiles and VPFIT solutions 
of the Mg II metal system at $z$=0.948.  The dashed 
vertical line corresponds to $z$=0.948}
\label{0948V}
\end{figure}

	The velocity plots and VPFIT solutions for the Mg II 
system at $z = 0.948$ are presented in Figure~\ref{0948V}.  The 
vertical lines above the Fe II 2382 and Mg II 2796 profiles indicate
the velocity centroids of the Gaussian components.
Fe II 2344 was excluded from the least-squares fit because of blending
with Mn II 2606 from the $z = 0.752$ system, while 
Fe II 2374 and Fe II 2586 were omitted given their
very low SNR.  The fit was reasonably successful yielding 
$\chi^2_\nu = 1.354$ with 3 components for
the Fe II transitions and 5 components for the Mg II transitions.
The solutions are listed in Table~\ref{0948F}.  Note the lack of Mg I
absorption at $v = 55$ \kms, possibly indicating a difference in the
ionization state of the two absorption features.

\subsubsection{$z$=1.017}

\begin{figure}
\centerline{
\psfig{figure=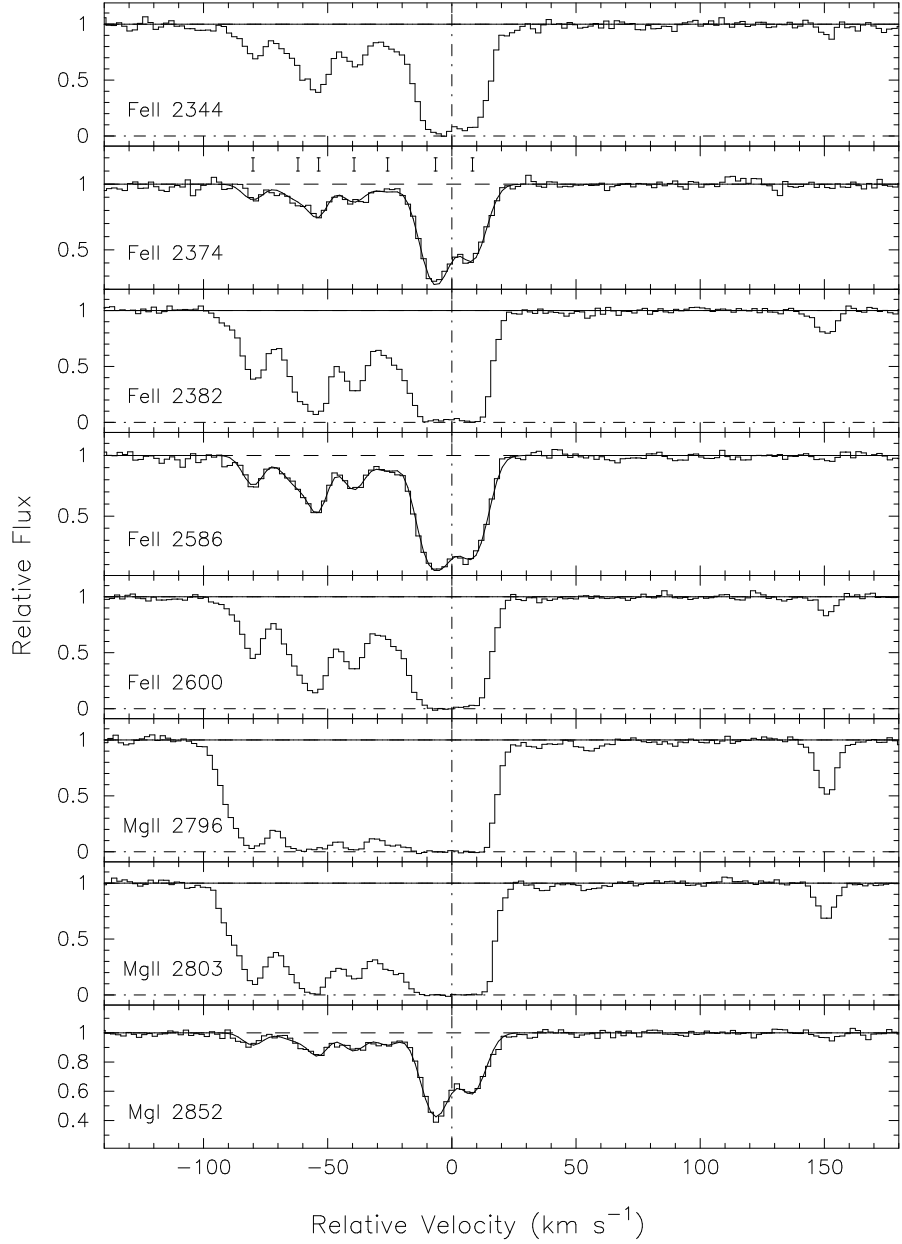,height=7.5in}}
\caption{ Velocity profiles and VPFIT solutions
of the Mg II metal system at $z$=1.017.  The dashed 
vertical line is at $z$=1.10172.}
\label{1017V}
\end{figure}

	Figure~\ref{1017V} shows the velocity profiles and VPFIT
solutions of the Mg II system at $z = 1.017$.   The velocity
centroids of the Gaussian components are given by the vertical
lines above the Fe II 2374 profile.
The weak feature at $v = 150$ \kms
is certainly real and may be considered associated with the
stronger absorption system.  
Given the high degree of line saturation, 
we were able to perform a successful fit
only to the unsaturated Fe II 2374, 2586 and Mg I 2852 profiles.  The
results of this fit are presented in Table~\ref{1017F}.  The solution
has a rather high reduced $\chi_\nu^2 = 1.36$ value primarily 
resulting from the Mg I solution.
Because the solution follows the Mg I profile very closely
(in contrast to the $z=0.948$ system), we suspect
we may have underestimated the error in the data in this region.

\subsection{Other Systems}

\subsubsection{$z = 2.014$}

\label{phosys}

\begin{figure}
\centerline{
\psfig{figure=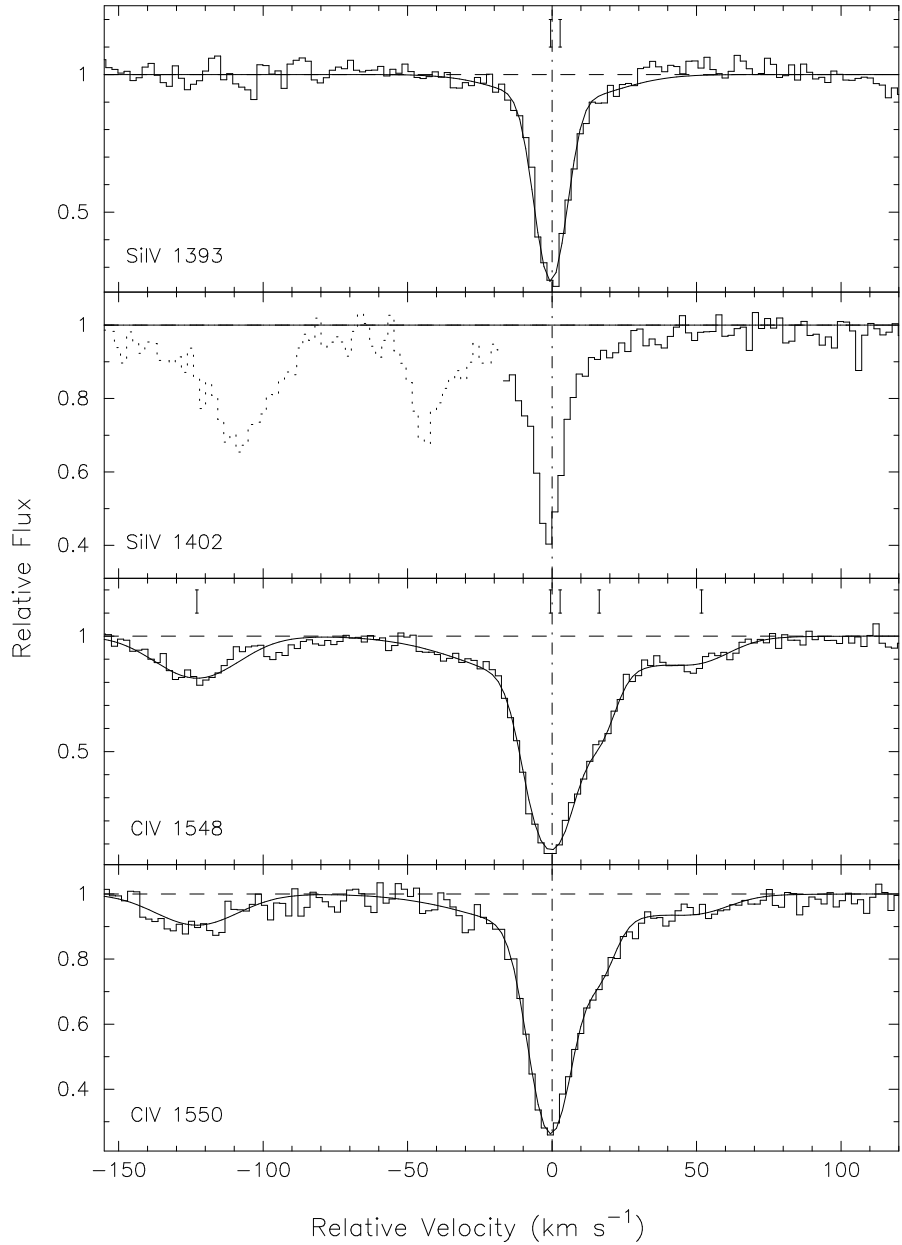,height=7.5in}}
\caption{ Velocity profiles and VPFIT solutions of the C IV system at 
$z$=2.014.  The high-ion profiles are very narrow, indicating
the system was not collisionally ionized.
The dashed vertical line corresponds to $z$=2.01438. }
\label{2014V}
\end{figure}

	The velocity profiles and VPFIT solutions for the 
high ions of the C IV system at $z = 2.014$ 
are plotted in Figure~\ref{2014V}.  
Si IV 1402 is blended with MgII 2803 from the $z=0.508$ system
and was not included in the fit.  The small vertical lines above
the Si IV 1393 and C IV 1548 transitions denote the velocity centroids
of the Gaussian components in the line-profile solution where we
have added 3 components to fit the C IV profiles.  
The VPFIT solution (see Table~\ref{201F}) yields an excellent
reduced $\chi_\nu^2 = 1.008$ with probability $P_{\chi^2} = 0.45$.

For this system we allowed for the possibility that the Doppler
broadening ($b$ values) of the features is not entirely due to
bulk motions.  Specifically we allowed for thermal broadening of the
$b$ values, which will yield different $b$
values for ions with different masses ($b \propto m^{-{1\over 2}}$) even
though the $b$ values for a given feature are tied together.
The solution suggests two components at $v \approx$ 0 \kms 
with very different physical properties.  
One is very narrow with $b$ values of
8.0 \kms and 5.2 \kms for C IV and Si IV respectively.  
Assuming these values to be thermal in origin, 
we find the temperature $T = 4.7 \sci{4} \; \rm K$. 
The other component is broad, with $b$ values of 38.7 \kms and
25.3 \kms ($T = 1 \sci{6} \; \rm K$) for C IV and Si IV respectively.

The ionic concentrations for C$^{+3}$ and Si$^{+3}$ 
formed through collisional ionization peak
at $T = 10^5 \; \rm K$ and $8\sci{4} \; \rm K$ 
respectively (\cite{cox,shl82}).
If collisional ionization were to explain the narrow component at
a temperature of $T \approx 5 \sci{4} \; \rm K$, 
we would expect to see significant concentrations of the singly
ionized species (C$^+$ and Si$^+$).  Unfortunately the 
C II 1334 transition is lost in the \lya forest, but we can perform an
accurate measurement of Si II 1526.  Measuring the equivalent
width over the velocity region spanning the Si IV feature, we find
a 3$\sigma$ upper limit to the Si$^+$ column density $\N{Si^+} < 12.43$,
which is considerably lower than the column density of Si IV.  Therefore
we contend the narrow component at $v \approx 0$ \kms does not 
result from collisional ionization.  
Furthermore, the simultaneous presence of low and high ions is a 
suggests the ions are produced by a photoionization flux.   
York et al. (1984) were the first to 
report a C IV system in which that ion could not have arisen from
collisional process, yet their results were tentative because of
limited resolution and signal to noise.  
Hence, the C IV system at $z = 2.014$
serves as the first unambiguous detection of a C IV
system formed through a process other than 
collisional ionization.

\subsubsection{$z = 2.281$ and $z=2.445$}

\begin{figure}
\centerline{
\psfig{figure=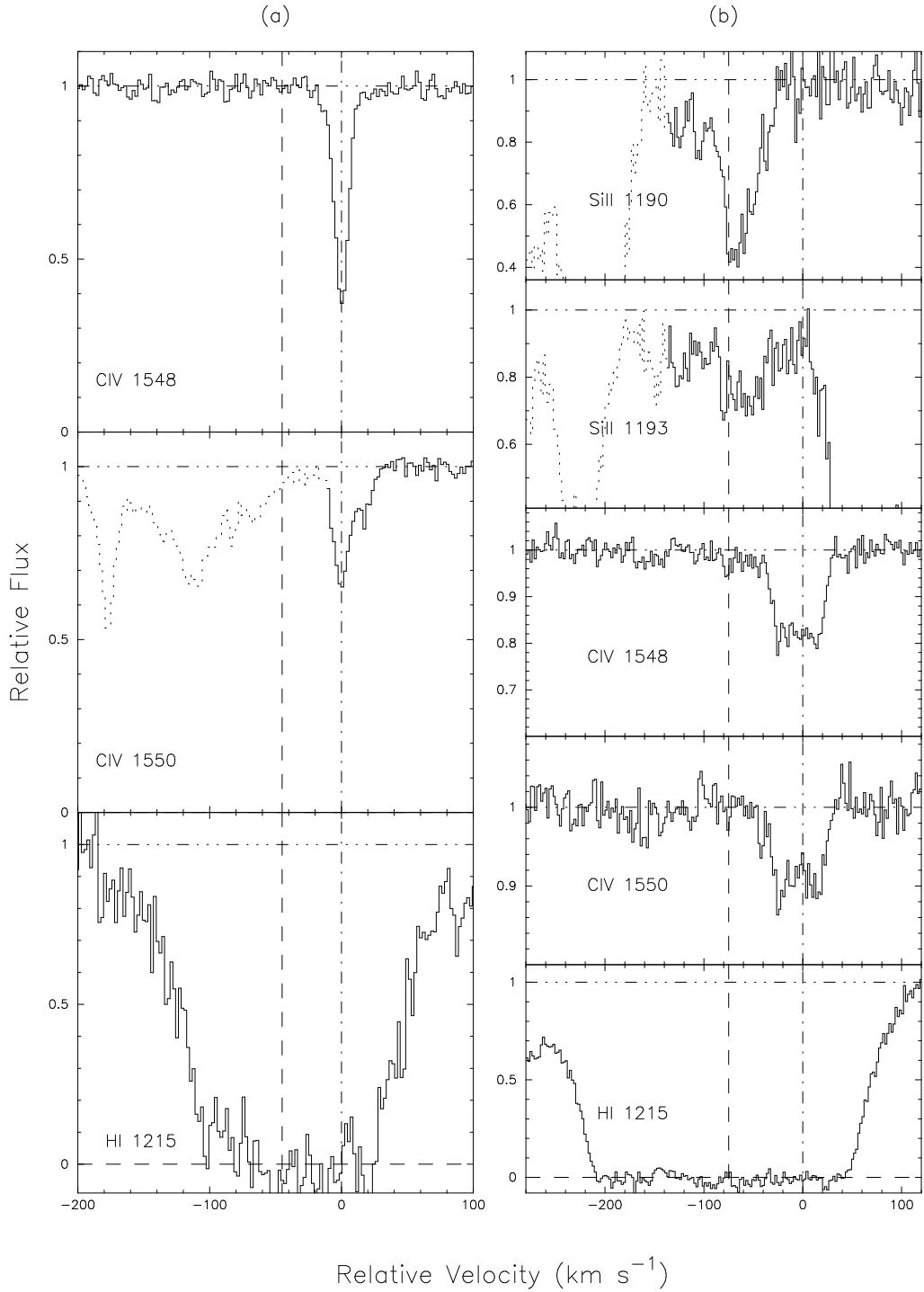,height=7.5in}}
\caption{ Velocity profiles of the C IV systems at 
(a) $z$=2.281 and (b) $z=2.445$.  
Note in each case the displacement of the center of the C IV
doublet from the center of the \Lya profile.  This is direct
evidence that high-ions do not trace HI gas.}
\label{2281V}
\end{figure}

	Figure~\ref{2281V}a is a velocity profile plot of the 
\lya and C IV transitions from the C IV system 
at $z = 2.281$.  Figure~\ref{2281V}b
shows Si II as well as the \Lya and C IV
transitions from the $z=2.445$ C IV system.  C IV 1550 at $z = 2.281$
is blended to the blue by Ni II 1751 from
the $z = 1.920$ system and to the red by an unidentified
transition.
The vertical dash-dot lines indicate the velocity
centroid of the C IV systems and the vertical dashed lines
mark the approximate velocity centroids of the \lya profiles.
The striking feature of these systems
is the obvious displacement ($\approx 40$ \kms for $z=2.281$ and
$\approx 100$ \kms for $z=2.445$) of 
C IV absorption from the center of the \Lya profile.  
In contrast,  the Si II profiles
from the $z=2.445$ system are more nearly centered
on the \Lya profile.
This clearly suggests C IV profiles do not generally
track HI or low-ion transitions.  

\section{IONIC COLUMN DENSITIES}
\label{ionden}

	This section presents the ionic column densities of several of
the metal line systems toward Q2206-199.  In many cases,
we perform hidden component analyses to investigate the 
presence of hidden saturation (\cite{sav91}).  
Both the line-profile fitting results and the apparent
optical depth method are used to derive column densities.  

The hidden component analysis and apparent optical depth method
may be summarized in the following manner.  One first calculates
the apparent ionic column density for each pixel, $N_a (v)$, 
according to the following equation:

\begin{equation}
N_a(v) = {m_e c \over \pi e^2} {\tau_a(v) \over f \lambda} ,
\end{equation}

\noindent where $\tau_a(v) = \ln [I_i (v) / I_a (v)]$, f is
the oscillator strength, $\lambda$ is the rest wavelength, and
$I_i$ and $I_a$ are the incident and measured intensity.
One then compares $N_a (v)$ profiles from multiple transitions of the same
ion to investigate line saturation.  Because $\tau_a (v)$ underestimates
the true optical depth of strong, unresolved transitions, the inferred
$N_a (v)$ will be smaller than for optically thin, weaker transitions.
This comparison analysis is termed a hidden component analysis.
Additionally, one may sum the $N_a (v)$ profile 
over velocity space to derive a total ionic column density. This 
procedure is termed the apparent optical depth method.

\subsection{Damped \Lya System at $z = 1.920$}

	In this section, we perform a comprehensive analysis of the 
ionic column densities for the damped \Lya system 
at $z = 1.920$.  We first 
perform a hidden component analysis to investigate 
the likelihood of hidden 
line saturation.  We then calculate the ionic column densities 
and 1$\sigma$
errors and finally make a quantitative evaluation of
the effect of hidden saturation in the Ni II and Cr II profiles.

\subsubsection{Hidden Component Analysis}
\label{shca}

\begin{figure}
\centerline{
\psfig{figure=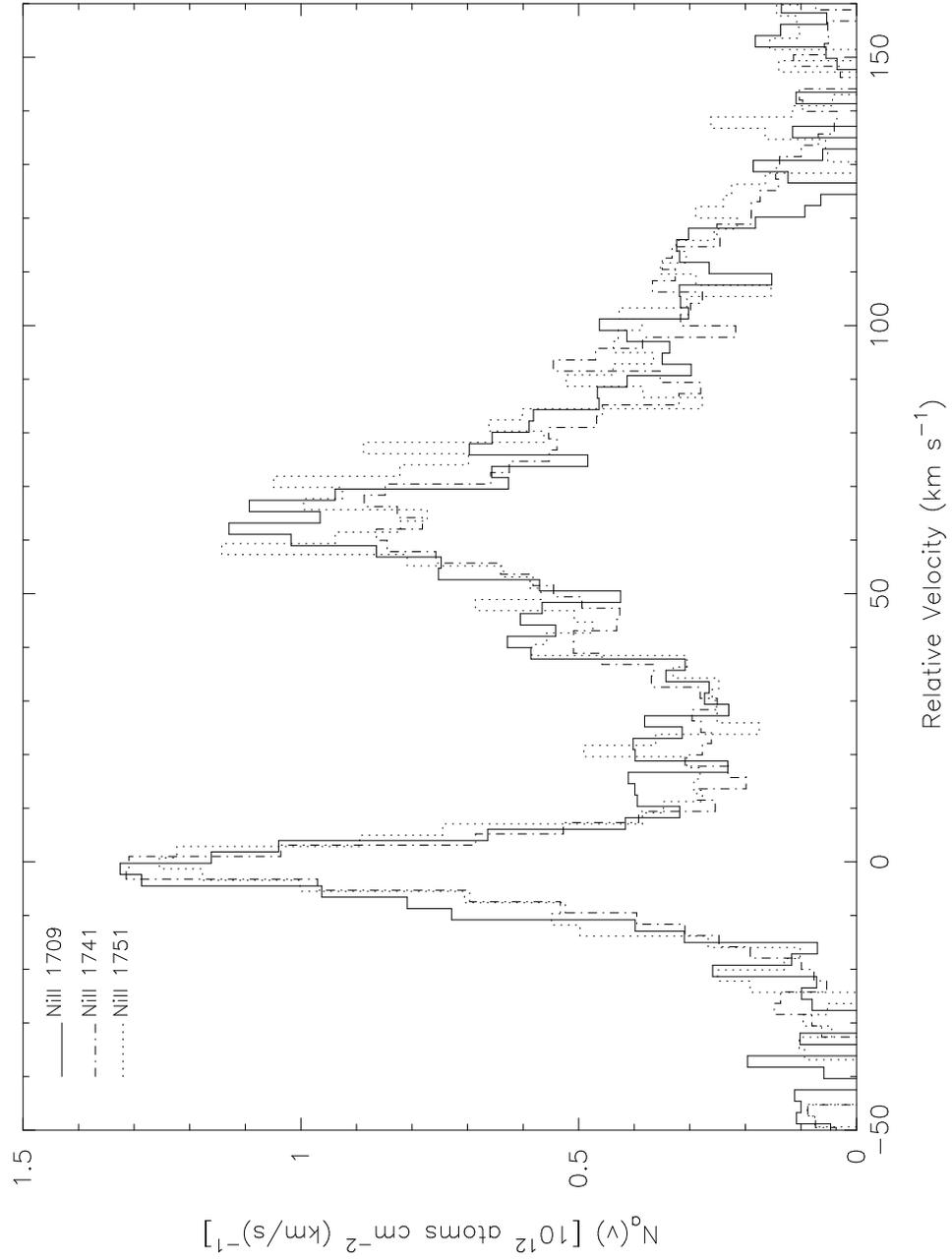,height=7.5in}}
\caption{ Apparent column density $[N_a (v)]$ profiles for Ni II 1709,
1741, and 1751 in the 
$z$=1.920 system.  Note that where the profile of the weakest
transition (Ni II 1709) dominates, this is evidence for hidden
saturation.  Therefore,
the divergence of the profiles near the features
at $v \approx 65$ \kms indicates hidden line saturation.} 
\label{1920Ni}
\end{figure}

	Figure~\ref{1920Ni} plots $N_a (v)$ for the 
Ni II 1709, 1741, 1751 
transitions of the damped \Lya system at $z = 1.920$.
The only feature where the $N_a (v)$ curves diverge substantially is at
$v = 65$ \kms.  Here the strong Ni II 1741 profile depth is significantly 
smaller than those of the other two weaker transitions.  
It is possible that there are at least two very narrow 
components $(b < 5$ \kms) in this region 
which are blended in the Ni II 1741 profile.
This is an example of hidden saturation, albeit a rather
minimal case.

\begin{figure}
\centerline{
\psfig{figure=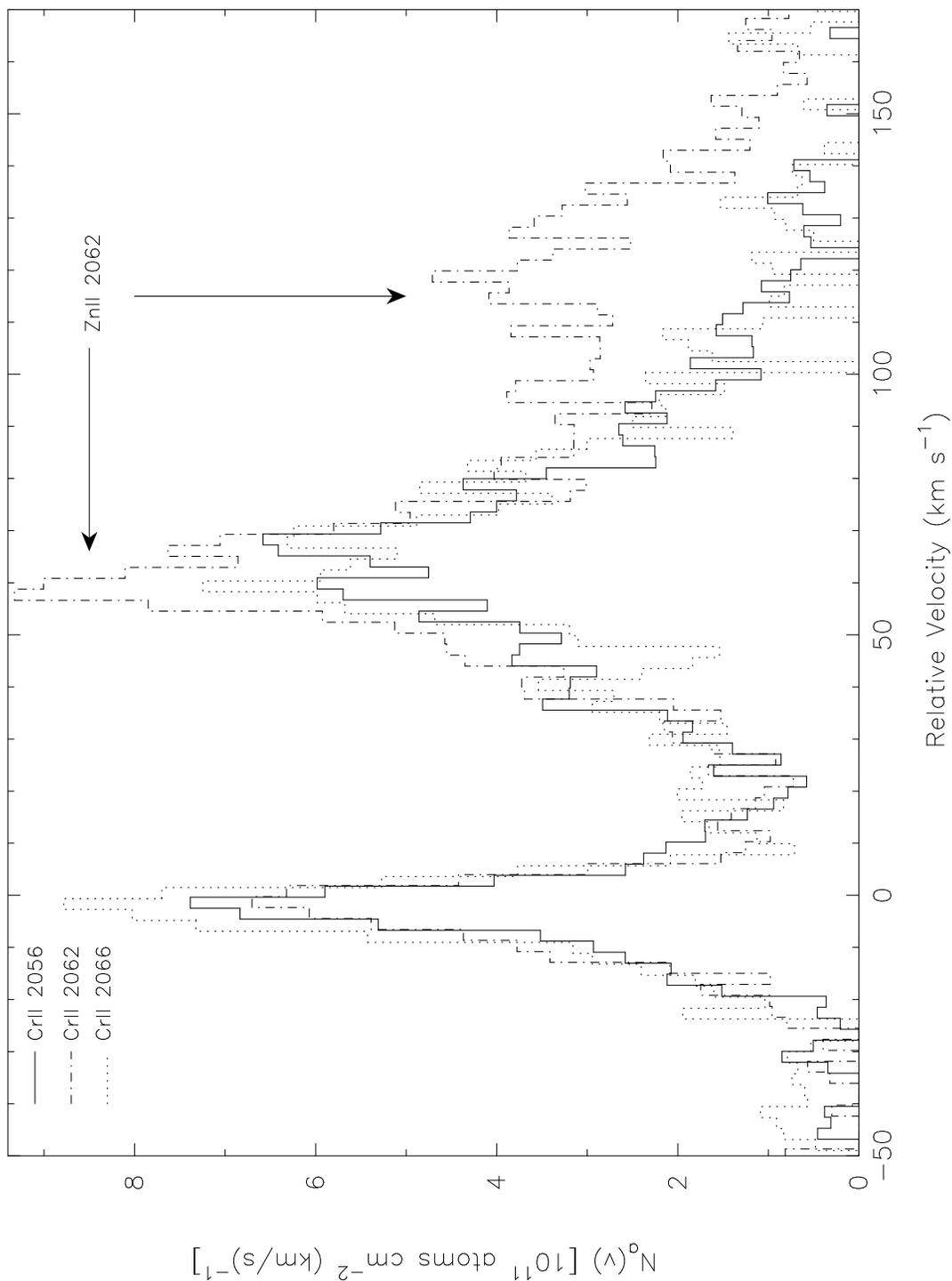,height=7.5in}}
\caption{ Apparent column density $[N_a (v)]$ profiles for Cr II 2056,
2062, and 2066 in the 
$z$=1.920 system.  Note that where the profile of the weakest
transition (Cr II 2066) dominates, this is evidence for hidden
saturation.  Therefore, the divergence of the profiles at $v = 3$ \kms
indicates hidden saturation.  The overabundance of Cr II 2062 at
$v > 40$ \kms is due to line blending with Zn II 2062.}
\label{1920Cr}
\end{figure}

	The hidden component analysis of the Cr II 2056, 2062, 2066 
transitions is presented in Figure~\ref{1920Cr}.  There is a clear
indication of hidden saturation in the bluest feature, 
where the weaker transition (Cr II 2066) dominates.
There is also a hint of hidden saturation 
in the feature at $v = 58$ \kms corresponding to 
the hidden saturation observed in the Ni II analysis.  
The dominant feature at this velocity, though,
is the blending apparent in the Cr II 2062 profile due to absorption from
Zn II 2062.  The divergence between the Cr II profiles
at $v \approx 130$ \kms is also due to absorption
from Zn II 2062.  Given the 
results of the two hidden component analyses, we consider the
low-ion profiles to be essentially free of hidden saturation 
with minor exceptions.   In order to quantify the effects of hidden
saturation on the ionic column densities of these two ions we
proceed to implement a correction technique described by Savage and
Sembach (1991).  

\subsubsection{Corrections for Hidden Saturation}

\label{hcacorr}

        As the hidden component analyses of Ni II and Cr II
suggest, there is minor hidden saturation evident
in several of the low-ion profiles.
In order to evaluate the error associated with this saturation, 
we use the correction technique proposed by Savage and Sembach (1991). 
We choose to focus on the Cr II saturation because the relative
oscillator strengths ($f_{2056} : f_{2066} \approx 2:1$) correspond
very closely to the analysis by Savage and Sembach (1991).
Adopting their notation, express the corrected column density $N_a^c$
as

\begin{equation}
\log N_a^c = \log N_a^{n-1} + \Delta \log N_a^{n-1}
\end{equation}

\noindent where $N_a^{n-1}$ is the column density of the weaker
transition, $\Delta \log N_a^{n-1}$ is derived numerically
from the difference between $\log N_a^{n-1}$ and $\log N_a^n$, and
$N_a^n$ is the column density of the stronger transition\footnote{
Hence $n$ denotes the multiple transitions of an ion with
increasing $n$ denoting larger $f$ values.}.  Using
their results (Table 4; \cite{sav91}) and our Cr II measurements
($\log N_a^{n-1} = 13.681$, $\log N_a^n = 13.628$), we find
$\Delta \log N_a^{n-1} = 0.053$ implying $\log N_a^c = 13.735$.
Thus the hidden saturation in Cr II may lead to an underestimate
of the true column density, but by only $\approx$ 0.05 dex. 
Note that the Ni II 1741, 1751 measurements ($\log N_a^{n-1} = 13.911$, 
$\log N_a^n = 13.859$, $\Delta \log N_a^{n-1} = 0.052$) 
give very similar results.   
This effect is on the order of the error associated with 
measuring the column densities 
and is therefore not important.  In short,
we do not expect abundances inferred from our ionic column densities
to suffer significantly from hidden saturation.

\subsubsection{Ionic Column Densities and Errors}

	The measured ionic column densities and 1$\sigma$ errors
for Fe$^+$, Ni$^+$, Si$^+$, Cr$^+$, and Zn$^+$ are listed in 
Table~\ref{1920I} as measured by both the 
apparent optical depth (column 2) 
and line profile fitting (column 3) methods.
For the line profile method, we 
summed the column densities of the individual components of each
transition and calculated the 1$\sigma$ error in the total value with 
standard least squares techniques.  The errors derived for the apparent
optical depth method are derived solely from Poisson statistics
and are systematically lower than the VPFIT solutions.  We believe
Poisson statistics alone underestimates
the 'true' error and therefore adopt the VPFIT error 
in all cases where
VPFIT solutions were obtained.  The adopted value for the ionic
column density is determined by averaging the measurements from the
two methods with the exception of Ni$^+$ and Cr$^+$ which have
been corrected by the techniques discussed in the previous subsection.
Note that in every case the two methods yield nearly the same value, 
further indicating a lack of significant hidden saturation.

	Table~\ref{1920I} also gives the ionic column densities for the
s-process transition Cu II 1358, e-process transition Ge II 1602, and
Ti II 1910 transitions.  
As these transitions are very weak, we use the linear curve of growth
to infer column densities from the measured rest-frame equivalent 
widths.  The rest frame equivalent widths are:
$\W{Cu^+} = 0.0526 \pm 0.0082$, $\W{Ge^+} = 0.0081 \pm 0.0028$, and
$\W{Ti^+} = 0.0348 \pm 0.0032$.  The ionic column densities for
Cu$^+$ and Ge$^+$ are derived trivially from the rest frame 
equivalent width.
The Ge$^+$ column density is reported as 3$\sigma$ upper limit.
The Cu II transition is detected at the 6$\sigma$ level, but this
may be an artifact of the very poor SNR at 3960$\rAA$.  The equivalent
width for the Ti$^+$, however, results from the
blend of the two transitions.  As the transitions are very weak,
we can derive a column density for Ti$^+$ assuming line
blending does not adversely effect the equivalent width  (i.e.
$W_{tot} = W^a_{1910} + W^b_{1910}$).  Using the $f$ values listed in
Table~\ref{osc}, we find $\log \N{Ti^+} = 12.807 \pm 0.038$.  This
represents the first confident ($> 5 \sigma$) detection of Ti
in a QSO absorption system.

\subsection{Damped \Lya System at $z = 2.076$}

	Unlike the damped system at $z = 1.920$, there are no multiple
transitions of any ion in the $z= 2.076$ damped \Lya system.  
Thus we cannot derive any information regarding 
hidden saturation.  However, given the very 
low metallicity of the system and
the correspondingly weak absorption observed,
we do not expect significant hidden saturation effects.

	Table~\ref{2076I} presents the ionic column densities of Si$^+$,
Fe$^+$, Al$^+$, Si$^{+3}$ and C$^{+3}$ as derived from profile
fitting and apparent optical depth methods.  Similar to 
the analysis of the $z = 1.920$ system,
we adopt a final value from a weighted average of the two measurements
and adopt the VPFIT error.
Also included in Table~\ref{2076I} are upper limit measurements
for the column density of Ni II 1751 (Ni II 1741 is between orders), 
Cr II 2056, and Zn II 2026 derived from both
the rest-frame equivalent width (assuming the linear curve of growth) and
the apparent optical depth method.  The measured rest-frame 
equivalent widths are:  $\W{Ni^+} = - 0.0009 \pm 0.0012$, 
$\W{Cr^+} = 0.0052 \pm 0.0014$ and $\W{Zn^+} = 0.0028 \pm 0.0013$.  
The negative
equivalent width value for Ni II 1751 is due to continuum error
and is consistent with a null detection.  
The column densities for Ni II and Zn II are reported as 
3$\sigma$ upper limits while Cr II 2056 was detected at the $3 \sigma$
level.  It is very possible, however, that the Cr II detection is
the result of error due to continuum fitting or poor
sky subtraction.

\subsection{Mg II Systems}

\subsubsection{$z = 0.752$}

\begin{figure}
\centerline{
\psfig{figure=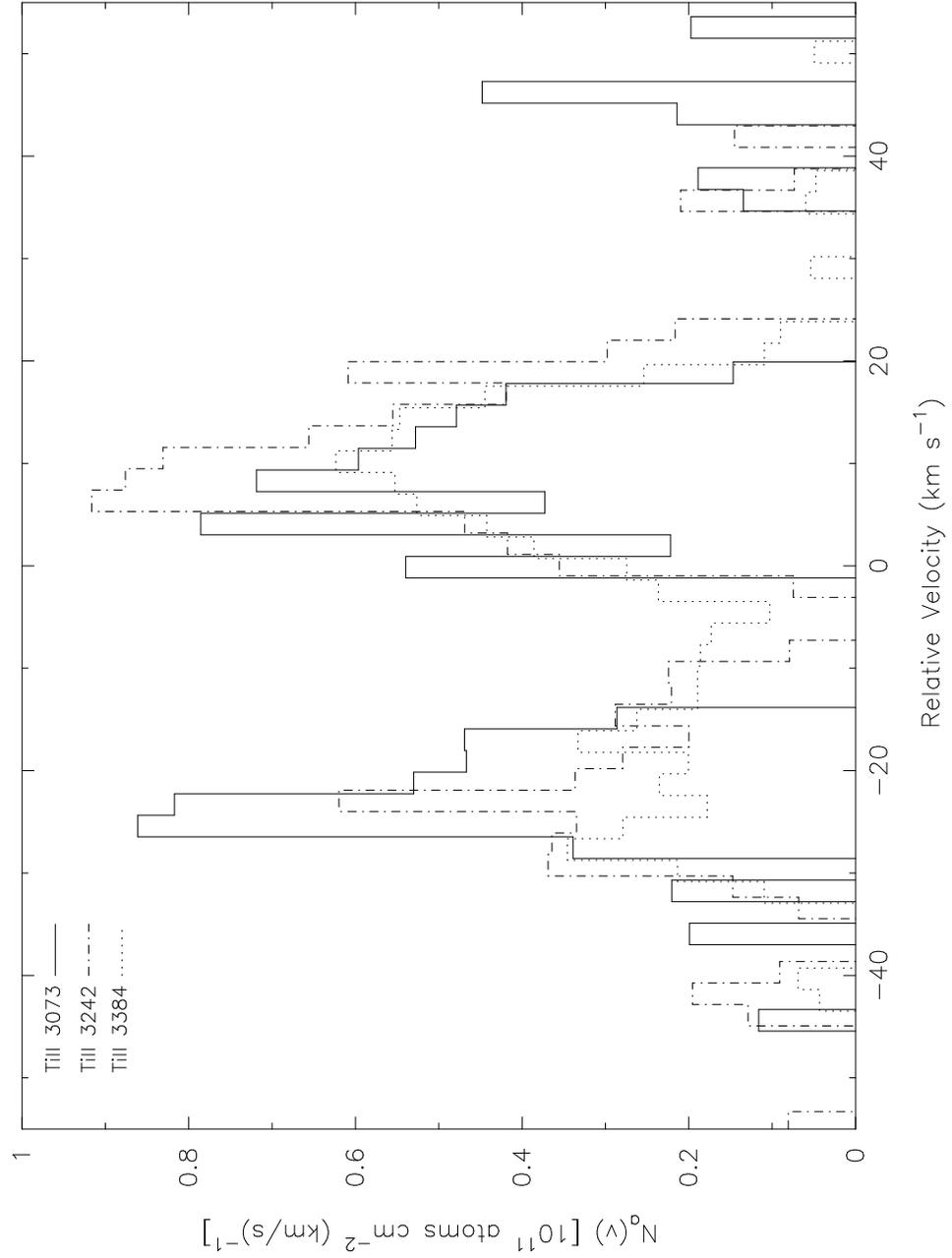,height=7.5in}}
\caption{ Apparent column density $[N_a (v)]$ profiles for Ti II 3073
3242, and 3384 in the 
$z$=0.752 system.  The divergence of the profiles at $v = -20$ \kms
may indicate hidden saturation.}
\label{hca_075}
\end{figure}

	Because of significant line saturation in the Mg II system at 
$z = 0.752$ (Figure~\ref{0752V}), the line profile fitting method
was unsuccessful.  Therefore, we obtained column densities for
this system from the profiles which 
are unsaturated only with the apparent
optical depth method.  These results are listed in Table~\ref{0752I}.
A discrepancy exists in the Mn II results (Mn II 2576
has a significantly higher value) which may
be due to errors in the oscillator strengths or possibly an 
unidentified blend.  Recalling Mn II 2606 is blended with
Fe II 2344 from the MgII system at $z = 0.948$, we adopt a final
value by averaging the column densities for Mn II 2576 and Mn II
2594.   Figure~\ref{hca_075}  is a plot of $\Nav$ for
the Ti II transitions.  Because these are very
weak transitions, the analysis is not sensitive, but
there is moderate evidence for hidden saturation at $-20$ \kms.
The values for the total column densities from the apparent optical
depth method all yield the same value within error and therefore
the effect of hidden saturation is minimal.  We adopt a Ti II column
density for this system by averaging the three values weighted
by their respective errors.

\subsubsection{$z = 0.948$}

	Table~\ref{0948I} lists the ionic column 
densities for the two clouds
associated with the Mg II system at $z = 0.948$.  There is a 
significant difference between the column density for Mg II as measured
by the two methods.  It is very likely hidden saturation may explain
the majority of this difference.
There is also a minor discrepancy between the column densities 
measured for the Fe II transitions which again may be due to hidden
saturation, but also could be a result of continuum error.
For both Fe II and Mg II we suggest adopting the VPFIT values as
the VPFIT package does correct for instrumental smearing.

\subsubsection{$z = 1.017$}

	The ionic column densities for Fe$^+$, Mg$^0$ and Mn$^+$ derived 
from the apparent optical depth and line profile fitting methods
are presented in Table~\ref{1017I}.  Similar to the $z = 0.948$ system,
there is some evidence for hidden saturation in the Fe II profiles
(i.e. the weaker Fe II 2374 transition has the highest apparent optical
depth ionic column density).  We choose to adopt the VPFIT values
for the same reasons as above.  Note that applying the correction
procedure outlined in $\S\ref{hcacorr}$ yields nearly the VPFIT values.

\section{ABUNDANCES}

	This section presents and discusses the abundance measurements
for the two damped \Lya systems based on the ionic column densities
of $\S\ref{ionden}$.  Given the very large neutral Hydrogen column densities
observed, we assume the ionized fraction of Hydrogen is
insignificant (\cite{pro96a}) and that the metals are 
predominantly singly ionized (\cite{pro96a,lu95a}).  Therefore
the abundances are derived from the ionic column densities of 
H$^0$, Ti$^+$, Fe$^+$, Si$^+$, 
Cr$^+$, Ni$^+$, and Zn$^+$ without ionization corrections.  

	As a qualitative check, we can compare the Al III 1854 
and low-ion line profiles, where a relative overabundance of
Al III may suggest an ionized region.  With the exception of the 
velocity region near $v \approx 25$ \kms, 
the profiles trace one another closely.
There is a possibility that the gas is in 
a higher ionization state at this
velocity, but lacking an unsaturated Al II profile it is
difficult to quantify the effect on the abundances.
As such, we choose to ignore ionization corrections since this
one velocity region would have at most only a minor effect on
our results. 

\subsection{Damped \Lya System at $z = 1.920$}

\subsubsection{Results}

\label{metll}

	The logarithmic
column density $\log[\N{X}]$, and the logarithmic
abundance of element X relative to Hydrogen 
normalized to solar abundances,
[X/H] $\equiv \log[\N{X}/\N{H}] - \log[\N{X}/\N{H}]_\odot$, for the
damped \Lya system at $z$=1.920 are listed in Table~\ref{1920A}.
The abundances for this system are derived 
assuming $\N{H} = 4.5 \pm 0.8 \sci{20} \cm2$ (\cite{ptt94}) 
and standard solar abundances (\cite{and89}).  Given the quality of
the Keck data, the error in $\N{H}$ dominates the error in the
abundances.  Therefore measurements of the relative abundances
of different elements will have substantially smaller errors
associated with them.  This system has the highest metallicity
that we have analyzed for a damped \Lya
system with [Zn/H] = $-0.374 \pm 0.097$ and [Fe/H] = $-0.705 \pm
0.097$.  We report both Zn and Fe as indicators of metallicity
given the results of recent studies (see \cite{lu96a,pro96a}).
Note that our value for Zn is consistent with the lower limit
presented by Pettini et al. (1994). 

\subsubsection{Relative Abundances and Dust}
\label{relabd}

\begin{figure}
\centerline{
\psfig{figure=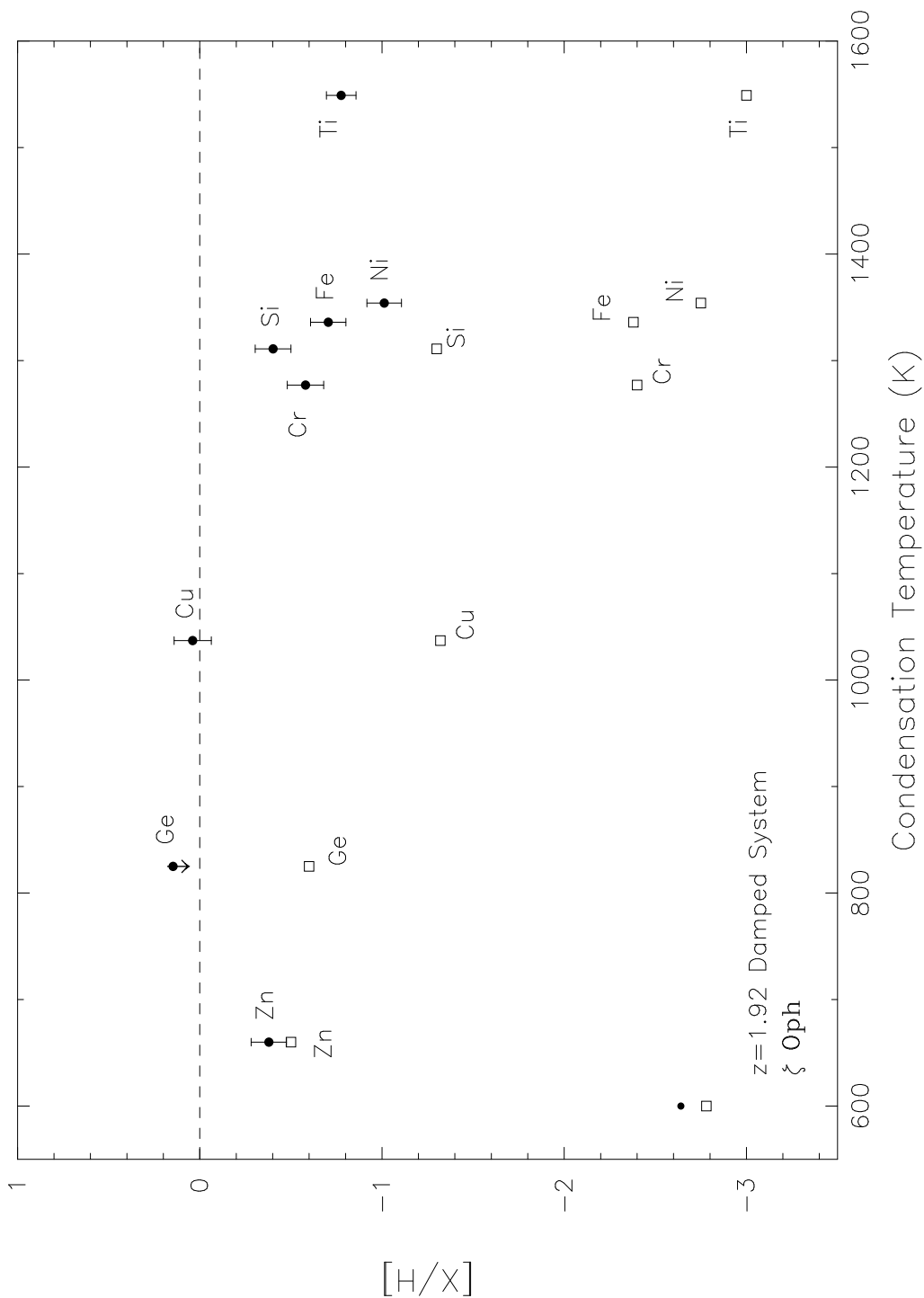,height=7.5in}}
\caption{ Logarithmic abundances of the low-ions 
Fe, Cr, Si, Ni, Zn, Cu and Ge relative to standard solar abundances
versus condensation
temperatures for the $z$=1.920 system (solid dots) and for the line
of sight in the ISM toward $\zeta$ Oph (open squares).  Error in the
neutral Hydrogen column density dominates the error associated with these
measurements such that the error in relative abundances of the
metals is significantly smaller. }
\label{1920T}
\end{figure}

Figure~\ref{1920T} plots the abundance
data versus condensation temperatures, $T_C$.  
Also shown in the figure are the corresponding values for
the sightline to $\zeta$ Oph in the ISM where the underabundance
of metals with respect solar abundances 
is assumed the result of depletion onto dust grains.
Comparing the damped \Lya
data with the ISM values reveals minimal evidence for a dust-like
pattern, in which Fe and Ni are underabundant relative to Zn. 
The equality of the Zn and Si abundance 
argues against this interpretation
since Si is underabundant relative to Zn in the Galaxy.  Moreover,
the abundance pattern we observe in the damped \lya data, in
particular [Ni/Fe], [Si/Fe] and [Cr/Fe], resembles the
pattern detected in halo stars in the Galaxy which are 
explained in terms of nucleosynthetic yields produced by type
II supernovae  (\cite{lu95b,lu96a,lu96b,pro96a}).  The Lu et
al.\ interpretation has one serious difficulty.  In all damped
\lya systems [Zn/Fe] $> 0$, whereas [Zn/Fe] $\approx 0$ for
stars with $-3 < $ [Fe/H] $< 0$ (\cite{sne91}).

Lu et al.\ (1996a) have stressed the importance of the relative
abundance of Mn to Fe as a strong indicator of the presence
of dust in damped \Lya systems.  For a large range of metallicities
[Mn/Fe] $> 0$ in the ISM, whereas Lu et al.\ (1996a) find
[Mn/Fe] $\leq 0$ for all of the damped  \lya systems they have 
analyzed.  In turn, they argue the abundance ratios of damped
\Lya systems are not consistent with dust depletion while
they are consistent with the yields from type II supernovae.  We can
apply similar arguments to the [Ti/Fe] ratio where depletion
in the ISM implies [Ti/Fe] $< 0$ over a wide range of metallicities.
For the damped \Lya system at $z=1.920$ we find [Ti/Fe] $= -0.071
\pm 0.069$, which is consistent (at the $1 \sigma$ level)
with no dust depletion.  Furthermore the
[Ti/Fe] ratio is consistent with the value observed in higher
metallicity Galactic stars (\cite{whe89}).

\subsubsection{Relative Abundances Variations and Dust}

\begin{figure}
\centerline{
\psfig{figure=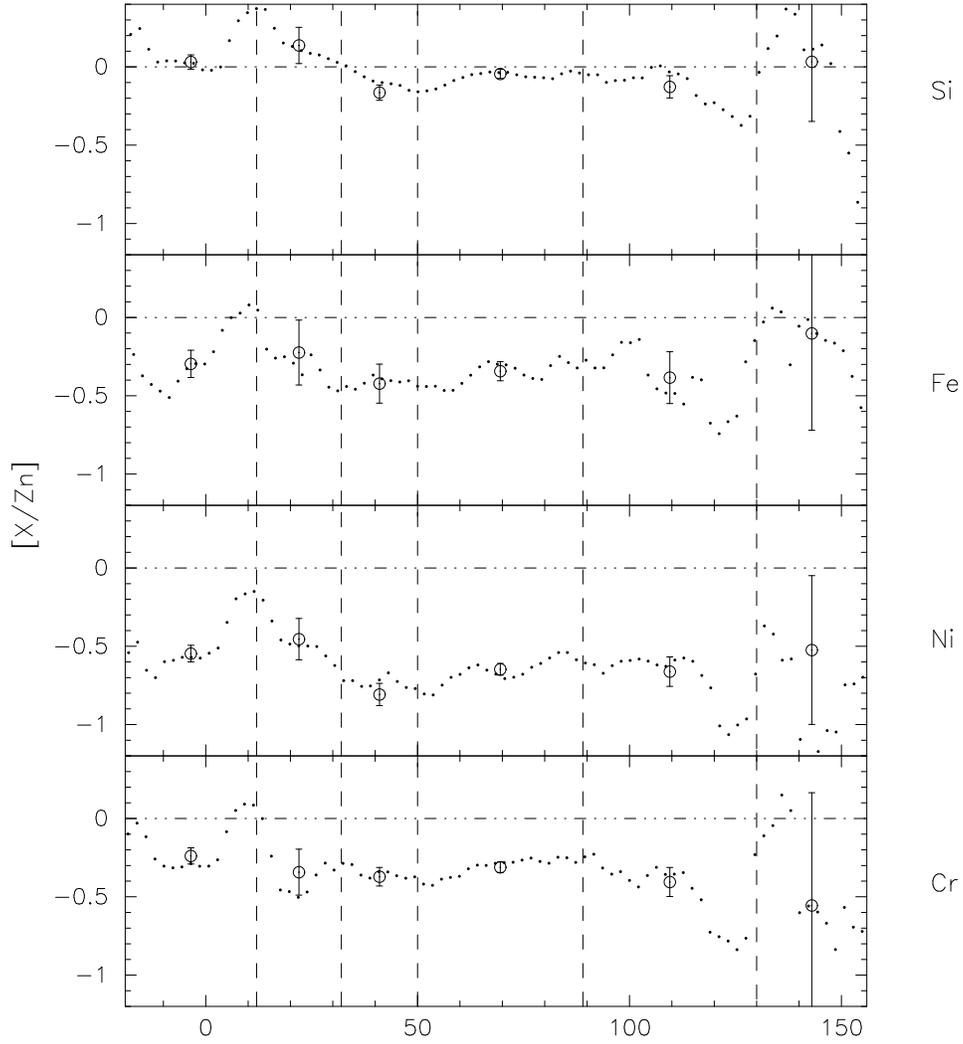,height=7.5in}}
\caption{ Relative cosmic abundances 
of Fe, Ni, Si, and Cr versus Zn for the $z = 1.920$ system.
The small points mark the relative abundance average over
5 pixels at the given velocity.  The large points denote
the relative abundance over the velocity intervals marked
by the large tick marks.  The dot-dash line at 0 
represents zero depletion with respect to Zn.}
\label{1920R}
\end{figure}

	Figure~\ref{1920R} is a plot of the abundance of Si, Fe,
Ni, and Cr relative to Zn in velocity space
assuming solar abundances.  The small
points represent a 5 pixel average of the relative abundance
at the given velocity, while the larger points represent relative
abundances over the velocity regions designated by the dashed vertical
lines.
There are several systematic features resulting from the shape
of the Zn II profile which need to be addressed.  First, the spike
at $v = 10$ \kms in each plot is most likely due to a sky emission
line that was poorly subtracted from the Zn II profile.  Secondly,
the slight dip at $v = 50$ \kms is due to a blend of Zn II 2026
with Mg I 2026.  Finally, the region at $v > 125$ \kms is dominated
by noise as all the profiles show essentially no absorption.

	Examining the averaged values, one notes the 
relative abundances remain
nearly constant over the entire velocity interval with only small
deviations primarily due to the systematic effects addressed above.  
Quantitatively,
the variation is on the order of $\approx 0.2$ dex.  
Using identical techniques to those presented in 
Prochaska and Wolfe 1996a,
we find variations of the hydrogen volume
density $n_H$ are no larger than $\approx 0.5$ dex if ISM-like dust
is present in this system.  
As discussed in Prochaska and Wolfe 1996a, we therefore
conclude either:
(1) there is an absence of grains with properties similar
to dust in the ISM or (2) the input of shocks driven by Supernovae in this
system is considerably weaker than that in the ISM.  
Furthermore, recent studies of metal abundances in the ISM
(\cite{spz93,spz95}) have indicated relative abundance (depletion)
variations of nearly 1.0 dex for Ni and Fe with respect to S.
Our data does not exhibit this same trend where we consider Zn,
a relatively undepleted element, instead of S.
Therefore, 
we contend the lack of abundance variations in this system suggests
the lack of a significant amount of ISM-like dust.

\subsection{Damped \Lya System at $z = 2.076$}

	Table~\ref{2076A} lists the column density and logarithmic
abundance of the metals associated with the damped \Lya system
at $z=2.076$.  The values for the Ni and Zn abundances represent $3 \sigma$
upper limits.  Because the Cr II transition is such a marginal
detection (i.e. the transition is dominated by noise), 
we believe its abundance may be significantly lower.  
In any case, this system is very metal
poor, exhibiting the lowest metallicity ([Zn/H] $< -1.745$, 
[Fe/H] $= -2.621 \pm 0.071$) 
we have measured for a damped system thus far.
The abundance measurements of Zn and Cr are a significant 
improvement over all previous work.  In the case of
Zn, for example, we have lowered the previous upper limit by nearly 1 dex. 

\begin{figure}
\centerline{
\psfig{figure=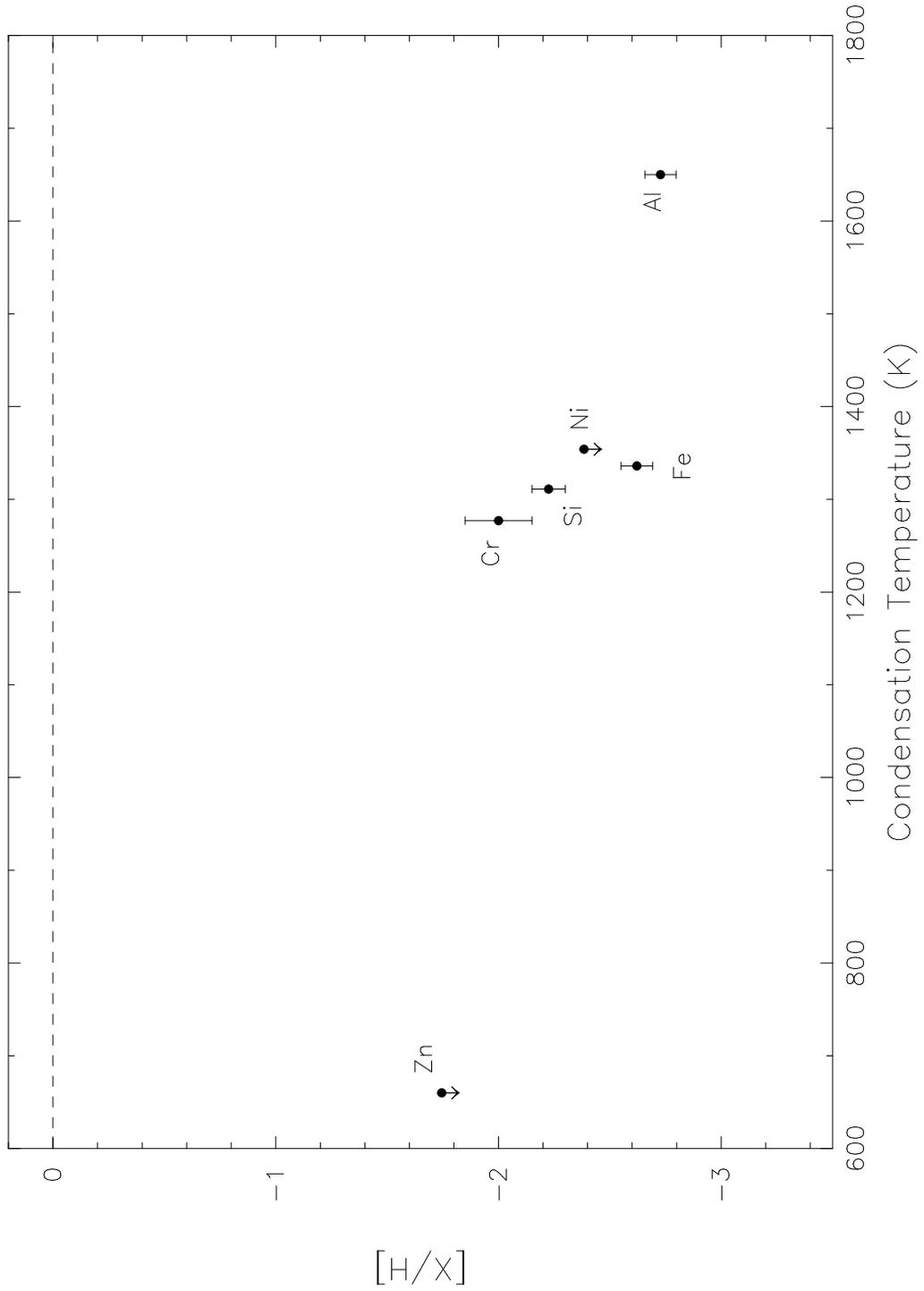,height=7.5in}}
\caption{ Abundances of the low-ions 
Fe, Cr, Si, Ni, Zn, and Al versus their condensation
temperatures for the $z$=2.076 system.  Note that with the
exception of Cr, the error in the
neutral Hydrogen column density dominates 
the error associated with these measurements.}
\label{2076T}
\end{figure}

	Figure~\ref{2076T} is a plot of the abundance data
as a function of condensation temperature. Note that
all of the abundances are consistent with an overall metallicity
[Zn/H] $< -1.745$.  Given the small number of metal transitions analyzed
here and the absence of a positive detection of Zn, 
it is very difficult to comment on the likelihood of dust
in this system.

\subsection{Damped? \Lya System at $z = 0.752$}

	If we assume solar abundances relative to Hydrogen
and assume the overall metallicity of this system is lower than
solar (i.e. [Zn/H], [Fe/H] $<$ 0.0 dex),
the Fe$^+$, Mn$^+$ and Ti$^+$ ionic column densities of the $z=0.752$
damped system imply 
$\log \N{HI} >$ 19.06, 19.00, and 19.37 respectively.  
Therefore, we believe it is very likely
that this system is in fact a damped \Lya system.  
As discussed in $\S\ref{relabd}$, the [Mn/Fe] ratio may be
a strong indicator of the presence (or lack thereof)
of dust grains in damped \Lya systems.  For this system, we find
[Mn/Fe] = $-0.099 \pm 0.061$ which further supports the results
of Lu et al.\ (1996a; see $\S\ref{relabd}$).  
Comparing the relative abundance of
Ti to Fe, we find [Ti/Fe] $= 0.269 \pm 0.042$ which also
is inconsistent with dust depletion, where [Ti/Fe] $< 0$ in
dust depleted regions within the ISM.  
Also, the [Ti/Fe] ratio is 
observed to increase with decreasing metallicity
in Galactic stars (\cite{whe89}), owing to the fact that 
Ti is an even-$Z$, $\alpha$-element.  Because the value of
[Ti/Fe] for the $z=0.752$ system is consistent with 
the values of Galactic halo stars,  we contend
this abundance ratio is consistent with the Lu et al.\ interpretation
and inconsistent with dust depletion.

\section{KINEMATICS}

\subsection{Damped \Lya System at $z = 1.920$}

     In this subsection we discuss the kinematics of the 
damped \Lya system at $z$=1.920. We stress the observed 
differences between the high and low-ion
profiles and compare the velocity profiles of the $z$=1.920 system 
with several other damped \Lya systems.

\subsubsection{Comparison of the Low and High Ion Profiles}

	Although the shape of the high-ion profiles (C IV and Si IV)
are dominated by line saturation, the profiles clearly differ from
the low-ion profiles.  For instance, the blue wing of the high-ion
profile is smoother and extends further than the low-ion blue wing.
In addition, there is significant absorption at 
several velocities in the high-ion
velocity profile (e.g. $v = 20$ \kms) not observed in the low-ion
transitions.  Finally, the velocity interval for the high ions
extends much further redward.  Given these observations, we contend
that the low and high ions are to a large degree kinematically 
disjoint.

	This characteristic has been observed in other damped \Lya
systems (\cite{wol94,pro96a}) we investigated with HIRES.  It is
very possible that the high-ion absorption may be from different
regions within the system.  Specifically, the high-ion absorption
may be due to hotter halo gas while the low ions may predominate
in the inner disk.  The large widths of the C IV profiles indicate
this may be the case.

\subsubsection{Comparison with Other Damped \Lya Systems}

	Our previous studies on the kinematics of damped \Lya
systems have revealed a systematic asymmetry (termed 
'edge-leading asymmetry') in the velocity
profiles.  In short, the low-ion profiles of the damped \Lya system
at $z = 2.309$ toward PHL 957 and at $z=2.462$ toward Q0201+365
have the strongest absorption feature at one edge of the velocity
interval and exhibit a decline in absorption toward the other
edge.  We believe this effect can be explained by passage of the
line of sight through a thick, rapidly rotating disk ($v_{rot} \sim 200$
\kms) in which the density of absorbing clouds decreases with radius and
height from midplane (\cite{wol95}).

	A similar effect (albeit less dramatic) is seen in the
low-ion profiles of the $z = 1.920$ damped \Lya system.  The
strongest feature lies at the blue edge of the velocity interval
with declining absorption toward the red.  By contrast with
the PHL 957 and Q0201+365 damped systems this system shows a 
lack of absorption over a significant velocity interval ($\Delta
v \approx 40$ \kms) between the two strongest features.
Beyond the second feature ($v > 50$ \kms) the profile
clearly possesses the edge-leading asymmetry.  We consider this
profile to be consistent with an edge-leading asymmetry, but note
that its shape is not as suggestive as the majority of the other
cases.
It is interesting to note that the Al III profile, which is not
identified as either a low or high ion, 
exhibits a very clear edge-leading
asymmetry.  

\subsection{Damped \Lya System at $z = 2.076$}

\subsubsection{Comparison of the Low and High Ion Profiles}

	Similar to the $z=1.920$ system, the high-ion profiles
of the $z=2.076$ damped \Lya system exhibit absorption over
over a wider velocity interval and at significantly different
velocities than the low-ion profiles.  However, unlike the $z=1.920$
system the Si IV and C IV profiles do not trace one
another very closely.  This may be predominantly due to
the lower metallicity or different ionization conditions.

\subsubsection{Comparison with Other Damped \Lya Systems}

	Given the very low metallicity and correspondingly
weak absorption features of the $z=2.076$ system, it is
difficult to compare its kinematics with other damped \Lya
systems.  The $z=2.076$ system does not show the edge-leading
asymmetry and is in fact a very symmetric profile.
In addition, the low and high-ion profiles span the
smallest velocity intervals observed thus far for a damped \Lya
system.  At the same time, however,
these characteristics are easily accommodated within
the thick rotating disk model.  The line-of-sight may be 
penetrating the disk nearly face-on or may be at several scale lengths
from the center of the disk thereby sampling a very small portion
of the circular velocity field.  
A future paper (\cite{pro96b}) will investigate
the likelihood of such an occurrence.

\section{SUMMARY AND CONCLUSIONS}

	We have presented high SNR HIRES spectra of the QSO Q2206-199
and have identified over $85\%$ of the features redward of \lya
emission (Table~\ref{abs}).  
We have analyzed 8 metal line systems in depth
and summarize the results as follows.

(1)  Although there is an indication of hidden line saturation
in the Ni II and Cr II transitions of the $z=1.920$ damped \lya system,
our analysis predicts the effect on the measured abundances 
is smaller than the 1$\sigma$ Poisson statistical errors.  Therefore
we confidently present abundance measurements of Fe, Cr, Ni,
Si, Ti and Zn relative to solar abundances
where the majority of error lies in the measurement of $\N{HI}$:
[Fe/H] = $-0.705 \pm 0.097$, [Cr/H] = $-0.580 \pm 0.100$,
[Si/H] = $-0.402 \pm 0.098$, [Ni/H] = $-1.012 \pm 0.095$, 
[Ti/H] = $-0.776 \pm 0.081$ and [Zn/H] = $-0.379 \pm 0.097$.  
The measurement of Ti is the first confident ($> 5 \sigma$)
detection in a QSO absorption line system.
In addition, we place a 3$\sigma$ upper limit on 
the Ge abundance ([Ge/H] $< 0.147$) and report a tentative detection
of [Cu/H] = 0.039 $\pm$ 0.102 in this system.

(2)  We have performed several analyses to test for the presence
of dust grains in the $z=1.920$ damped \Lya system.  Plots of
abundances relative to solar versus condensation temperature
exhibit little evidence for the presence of dust grains when
compared to similar plots for the sightline to $\zeta$ Oph.
The abundance pattern
of the damped system does follow the $\zeta$ Oph data in that 
Zn is overabundant with respect to Ni, Fe and possibly Cr, but
the differences are comparatively small.  In addition, [Si/H] $\approx$
[Zn/H] in the damped system which clearly contradicts the ISM data.
We contend this analysis is consistent both with the existence of ISM-like
dust at a level significantly below that of the sightline to $\zeta$
Oph as well as the abundance patterns of halo stars which 
have been enriched by type II supernovae.   Furthermore,
we measure [Ti/Fe] $\approx 0$ which is consistent with
no dust depletion.

We have
searched for abundance variations of Fe, Si, Cr and Ni relative to Zn  
over the velocity interval spanning the low-ion profiles.
We observe very small variations ($\approx 0.2$ dex) and have 
used techniques developed in previous work (\cite{pro96a}) to
conclude:  (a) either the input of shocks driven by supernovae
is considerably weaker in this system than that in the ISM or
(b) there is a very low level of ISM-like dust grains present
in the system.

(3)  The abundances of the damped \Lya
system at $z=2.076$ are very low.  Analyzing the strong Fe II, Si II
and Al II transitions we derived the following abundances:
[Fe/H] = $-2.621 \pm 0.071$, 
[Si/H] = $-2.225 \pm 0.075$,
[Al/H] = $-2.727 \pm 0.070$.
Even with our excellent
resolution and moderately high SNR we could only place upper limits
on the abundance of Zn and Ni ([Zn/H] $< -1.745$ and
[Ni/H] $< -2.384$).  We report a tentative measurement of Cr,
[Cr/H] = $-2.00 \pm 0.15$, whose value we question as noise
dominates the absorption profile. 
Lacking abundance data for Zn, we could not perform a detailed dust
analysis for this system.

(4) In addition to having widely different metallicities, the
$z=2.076$ and $z=1.920$ damped \Lya systems exhibit remarkably 
different kinematic characteristics.   The low-ion profiles 
(expected to trace the kinematics of these predominantly neutral systems)
of the $z=1.920$ system span $\approx 150$ \kms in velocity space
and are largely asymmetric.  Conversely, the low-ion profiles
of the $z=2.076$ system span only 30 \kms and are highly symmetric.
Even given these differences, 
the two profiles are consistent with the kinematics expected from
a thick, fast ($\approx 200$ \kms) rotating disk.  The $z=1.920$ 
profiles can be explained with a moderately inclined disk,
while the $z=2.076$ profile would be due to a nearly face-on
disk (low probability) or a line of sight which penetrated the
outer regions of the disk (more likely) where the radius of
curvature is small.  These ideas will be more fully developed in
future work (\cite{wol96,pro96b}).

(5)  We have investigated a very strong Mg II system
at $z=0.752$.  This system shows absorption in the very weak
Mn II 2606, 2594, 2576 and Ti II 3073, 3342, and 3384 transitions.
Together with the damped \lya system at $z=1.920$, 
this marks the first time Ti has been 
confidently observed in the IGM.  The column densities we derived from
these transitions imply a very conservative lower limit to
$\N{HI}$ of $\log \N{HI} > 19.00$, having assumed solar abundances
and no depletion.  We believe it is very possible that this is a 
third damped system toward Q2206$-$199.  

The relative abundance ratios of Mn and Ti to Fe strongly suggest
the system does not suffer from dust depletion.  Specifically,
[Mn/Fe] $= -0.099 \pm 0.061$ and [Ti/Fe] = $0.269 \pm 0.042$,
whereas in the ISM gas affected by dust depletion
has [Mn/Fe] $> 0$ and [Ti/Fe] $< 0$.  These results essentially
rule out the possibility of dust depletion and further
support the hypothesis of Lu et al.\ (1996a) that the abundance
patterns of damped \Lya systems are characteristic of type
II supernovae. 

(6)  We have identified a C IV system at $z=2.014$ with a very narrow
feature seen in C IV and Si IV absorption. 
The $b$ values (8.9 \kms and 5.2 \kms for C IV and Si IV
respectively) suggest a temperature of $4.7 \sci{4} \; \rm K$.
As collisional ionization dominates the production of these ions
at  $T > 8 \sci{4} \; \rm K$, we believe collisional ionization is
not the dominant process of formation.
This marks the first unambiguous evidence for
earlier claims (\cite{yor84}) of the existence of C IV systems
in the IGM formed by a process other than collisional ionization.

\acknowledgments

The authors would like to thank Bob Carswell for providing 
the line-profile
fitting package VPFIT as well as Tom Barlow for 
his excellent HIRES data 
reduction software.  AMW and JXP were partially 
supported by NASA grant NAGW-2119 and NSF grant AST 86-9420443.  

\clearpage

\begin{table*}
\begin{center}
\begin{tabular}{lccc}
UT Date
& Exposure
& Wavelength
& Resolution\\ 
& Time (s) & Coverage ($\rm \rAA$) & (\kms) \\
\tableline
1994 Sep 15 & 10800 & 3940 - 6360 & 7.2 - 8.0 \cr
1994 Sep 30 & 7200 & 4140 - 6520 & 7.4 - 8.0 \cr
1994 Oct 1 & 7900 & 4140 - 6520 & 7.2 - 8.0 \cr
\end{tabular}
\end{center}

\caption{JOURNAL OF OBSERVATIONS} \label{jou}

\end{table*}

\begin{table*}
\begin{center}
\begin{tabular}{lcc}
Order & $\lambda_{\rm center} \; \rAA$ & SNR \\
\tableline
90 & 3960 & 9.7 \cr
85 & 4190 & 21.8 \cr
80 & 4450 & 38.7 \cr
75 & 4750 & 34.7 \cr
70 & 5090 & 45.4 \cr
65 & 5480 & 60.8 \cr
60 & 5940 & 47.1 \cr
55 & 6480 & 41.4 \cr
\end{tabular}
\end{center}

\caption{SNR FOR SEVERAL ORDERS} \label{snr}

\end{table*}

\clearpage

\begin{deluxetable}{lcccll}
\tablewidth{0pc}
\tablenum{3}
\tablecaption{ABSORPTION LINE LIST}
\tablehead{
\colhead{Order} & \colhead{$\lambda$} &
\colhead{W} & \colhead{$\sigma_W$} &
\colhead{ID} & \colhead{$z_{abs}$} \nl
& \colhead{$(\rAA)$} &
\colhead{$(\rAA)$}}

\tiny
\startdata
90  \nl
& 3946.45 & 0.0651 &  0.0118 &           &       \nl
& 3950.40 & 0.2511 &  0.0156 &           &       \nl
& 3957.38 & 1.7486 &  0.0245 &           &       \nl
& 3960.83 & 0.0480 &  0.0085 & FeII 2260   & 0.752 \nl
& 3964.61 & 0.5092 &  0.0151 &           &       \nl
& 3968.53 & 0.0468 &  0.0082 &           &       \nl
& 3969.11 & 0.0537 &  0.0094 &           &       \nl
& 3972.20 & 0.3433 &  0.0139 &           &       \nl
& 3974.11 & 1.6655 &  0.0207 &           &       \nl
& 3977.75 & 1.2153 &  0.0228 &           &       \nl
& 3982.86 & 0.0818 &  0.0127 &           &       \nl
& 3984.10 & 0.5275 &  0.0271 &           &       \nl
89  \nl
& 3977.82 & 1.1607 &  0.0282 &           &       \nl
& 3984.39 & 0.4686 &  0.0222 &           &       \nl
& 3988.39 & 2.5651 &  0.0301 & Ly-a 1215   & 2.281 \nl
& 3992.31 & 1.4807 &  0.0268 &           &       \nl
& 3995.72 & 1.4815 &  0.0186 &           &       \nl
& 3998.46 & 0.3567 &  0.0142 &           &       \nl
& 4001.73 & 0.4066 &  0.0179 & NiII 1370   & 1.920 \nl
& 4005.76 & 0.2964 &  0.0084 & OI 1302     & 2.076 \nl
& 4008.02 & 1.7333 &  0.0231 &           &       \nl
& 4012.52 & 0.1380 &  0.0079 & SiII 1304   & 2.076 \nl
& 4015.99 & 0.1495 &  0.0136 &           &       \nl
& 4020.97 & 0.0760 &  0.0129 &           &       \nl
& 4022.39 & 0.1318 &  0.0134 &           &       \nl
88  \nl
& 4022.01 & 0.0957 &  0.0160 &           &       \nl
& 4037.14 & 2.3788 &  0.0217 & Ly-a 1215   & 2.321 \nl
& 4045.66 & 0.5778 &  0.0119 &           &       \nl
& 4047.16 & 0.0411 &  0.0081 &           &       \nl
& 4048.43 & 0.0628 &  0.0091 &           &       \nl
& 4049.37 & 0.3253 &  0.0113 &           &       \nl
& 4052.17 & 0.1348 &  0.0113 &           &       \nl
& 4057.66 & 0.9004 &  0.0168 &           &       \nl
& 4060.77 & 0.8117 &  0.0201 &           &       \nl
& 4068.01 & 0.0738 &  0.0107 &           &       \nl
& 4068.55 & 0.1006 &  0.0098 &           &       \nl
& 4070.42 & 2.5209 &  0.0229 & SiIV 1393   & 1.920 \nl
& 4073.78 & 2.3038 &  0.0268 &           &       \nl
& 4075.73 & 0.0634 &  0.0089 &           &       \nl
87  \nl
& 4068.33 & 0.0995 &  0.0120 &           &       \nl
& 4070.42 & 2.5209 &  0.0229 & SiIV 1393   & 1.920 \nl
& 4073.78 & 2.3038 &  0.0268 &           &       \nl
& 4087.44 & 0.9204 &  0.0172 &           &       \nl
& 4088.90 & 0.0378 &  0.0064 &           &       \nl
& 4089.96 & 0.1436 &  0.0110 &           &       \nl
& 4091.22 & 0.3615 &  0.0143 &           &       \nl
& 4094.65 & 0.0828 &  0.0101 &           &       \nl
& 4097.09 & 2.6619 &  0.0196 & SiIV 1402   & 1.920 \nl
& 4100.06 & 0.4152 &  0.0140 &           &       \nl
& 4105.10 & 0.5883 &  0.0150 & CII 1334    & 2.076 \nl
& 4106.79 & 0.8740 &  0.0113 & FeII 2344   & 0.752 \nl
& 4108.05 & 0.5139 &  0.0117 &           &       \nl
& 4110.05 & 0.3905 &  0.0181 &           &       \nl
& 4113.05 & 2.6078 &  0.0222 &           &       \nl
& 4115.74 & 0.2083 &  0.0131 &           &       \nl
& 4118.29 & 0.0592 &  0.0085 &           &       \nl
& 4120.18 & 0.4861 &  0.0139 &           &       \nl
86  \nl
& 4113.64 & 2.0268 &  0.0250 &           &       \nl
& 4115.84 & 0.1844 &  0.0158 &           &       \nl
& 4120.22 & 0.4658 &  0.0143 &           &       \nl
& 4123.89 & 2.3905 &  0.0204 &           &       \nl
& 4125.89 & 0.1217 &  0.0124 &           &       \nl
& 4127.63 & 0.0752 &  0.0102 &           &       \nl
& 4129.25 & 0.3551 &  0.0115 &           &       \nl
& 4132.64 & 0.1138 &  0.0081 &           &       \nl
& 4135.10 & 0.0625 &  0.0067 &           &       \nl
& 4143.38 & 0.4809 &  0.0087 &           &       \nl
& 4145.27 & 0.0220 &  0.0037 &           &       \nl
& 4152.07 & 0.1950 &  0.0065 &           &       \nl
& 4154.18 & 0.9786 &  0.0099 &           &       \nl
& 4159.77 & 0.6149 &  0.0078 & FeII 2374   & 0.752 \nl
& 4165.60 & 0.4197 &  0.0113 &           &       \nl
85  \nl
& 4165.69 & 0.4364 &  0.0132 &           &       \nl
& 4172.63 & 0.1147 &  0.0071 &           &       \nl
& 4174.51 & 1.2104 &  0.0094 & FeII 2382   & 0.752 \nl
& 4182.16 & 0.0274 &  0.0038 &           &       \nl
& 4186.41 & 4.7572 &  0.0128 & Ly-a 1215   & 2.445 \nl
& 4192.97 & 0.0689 &  0.0054 &           &       \nl
& 4197.51 & 0.1836 &  0.0058 &           &       \nl
& 4201.35 & 0.1796 &  0.0046 & SiIV 1393   & 2.014 \nl
& 4202.99 & 0.0152 &  0.0030 &           &       \nl
& 4206.27 & 0.0555 &  0.0050 &           &       \nl
& 4206.86 & 0.0313 &  0.0040 &           &       \nl
& 4208.68 & 1.2575 &  0.0072 &           &       \nl
& 4217.03 & 0.0666 &  0.0040 & MgII 2796 & 0.508 \nl
& 4218.43 & 0.0296 &  0.0049 &           &       \nl
& 4219.87 & 0.0316 &  0.0053 &           &       \nl
84  \nl
& 4217.04 & 0.0715 &  0.0041 & MgII 2796 & 0.508 \nl
& 4218.41 & 0.0659 &  0.0062 &           &       \nl
& 4219.80 & 0.0537 &  0.0059 &           &       \nl
& 4220.69 & 0.2278 &  0.0063 &           &       \nl
& 4226.86 & 0.1417 &  0.0053 &           &       \nl
& 4227.96 & 0.0717 &  0.0034 & MgII 2803 & 0.508 \nl
& 4228.49 & 0.1481 &  0.0044 & SiIV 1402   & 2.014 \nl
& 4231.22 & 0.3129 &  0.0068 &           &       \nl
& 4232.97 & 0.0416 &  0.0046 &           &       \nl
& 4234.16 & 0.1104 &  0.0051 &           &       \nl
& 4237.71 & 0.1946 &  0.0056 &           &       \nl
& 4240.00 & 0.2172 &  0.0055 &           &       \nl
& 4242.73 & 0.1727 &  0.0054 &           &       \nl
& 4245.22 & 0.1196 &  0.0046 &           &       \nl
& 4246.34 & 0.1254 &  0.0046 &           &       \nl
& 4247.80 & 0.1523 &  0.0044 &           &       \nl
& 4248.48 & 0.0098 &  0.0018 &           &       \nl
& 4249.22 & 0.1844 &  0.0052 & NI 1199   & 2.542 \nl
& 4252.45 & B\tablenotemark{a}  & & NiII 1454   & 1.920 \nl
& 4256.16 & 2.1368 &  0.0095 &           &       \nl
& 4260.37 & 0.0249 &  0.0031 &           &       \nl
& 4261.68 & 0.8938 &  0.0059 &           &       \nl
& 4267.67 & 0.0890 &  0.0055 & NV 1238     & 2.445 \nl
83  \nl
& 4261.74 & 0.9095 &  0.0070 &           &       \nl
& 4267.56 & 0.0920 &  0.0052 & NV 1238     & 2.445 \nl
& 4269.43 & 0.0421 &  0.0047 &           &       \nl
& 4272.22 & 0.1943 &  0.0048 & OI 1302?    & 2.281 \nl
& 4273.66 & 0.0156 &  0.0026 &           &       \nl
& 4274.53 & 0.0237 &  0.0028 &           &       \nl
& 4276.33 & 1.4103 &  0.0074 &           &       \nl
& 4279.40 & 0.1920 &  0.0046 &           &       \nl
& 4285.82 & B\tablenotemark{a}  & & NiII 1467   & 1.920 \nl
& 4288.65 & B\tablenotemark{a}  & & Si IV 1392  & 2.076 \nl
& 4293.15 & 0.0666 &  0.0033 &           &       \nl
& 4296.63 & 2.6738 &  0.0076 &           &       \nl
& 4300.57 & 0.1345 &  0.0040 &           &       \nl
& 4302.55 & 0.1181 &  0.0038 &           &       \nl
& 4306.60 & 2.6804 &  0.0051 & Ly-a 1215   & 2.542 \nl
& 4315.31 & 0.0850 &  0.0033 & SiIV 1402   & 2.076 \nl
& 4319.76 & 0.5776 &  0.0046 &           &       \nl
82  \nl
& 4315.28 & 0.0958 &  0.0042 & SiIV 1402   & 2.076 \nl
& 4319.81 & 0.5773 &  0.0047 &           &       \nl
& 4320.90 & 0.0156 &  0.0026 &           &       \nl
& 4322.33 & 0.1549 &  0.0041 &           &       \nl
& 4327.50 & 0.3559 &  0.0045 &           &       \nl
& 4329.53 & 0.0106 &  0.0020 &           &       \nl
& 4331.57 & 0.5283 &  0.0050 &           &       \nl
& 4334.76 & 0.2405 &  0.0038 &           &       \nl
& 4357.98 & 0.0823 &  0.0022 & SiIV 1393   & 2.126 \nl
& 4359.99 & 0.0520 &  0.0027 & SiIV 1393   & 2.128 \nl
81  \nl
& 4386.17 & 0.0617 &  0.0026 & SiIV 1402   & 2.126 \nl
& 4388.29 & 0.0260 &  0.0025 & SiIV 1402   & 2.128 \nl
80  \nl
& 4458.75 & 2.8803 &  0.0053 & SiII 1526   & 1.920 \nl
79  \nl
& 4512.28 & 0.0443 &  0.0029 &           &       \nl
& 4514.45 & 0.1441 &  0.0052 & MnII 2576   & 0.752 \nl
& 4521.84 & 3.3834 &  0.0084 & C IV 1548   & 1.920 \nl
& 4529.77 & B\tablenotemark{a}  & & C IV 1550   & 1.920 \nl
& 4538.29 & 0.0224 &  0.0033 &           &       \nl
78  \nl
& 4544.98 & 0.0642 &  0.0053 & MnII 2594   & 0.752 \nl
& 4555.25 & 1.1287 &  0.0051 & FeII 2600   & 0.752 \nl
& 4560.36 & 0.0174 &  0.0025 & FeII 2260   & 1.017 \nl
& 4566.44 & B\tablenotemark{a}  & & MnII 2606   & 0.752 \nl
& 4572.29 & S\tablenotemark{b}  & & SKY         &       \nl
77  \nl
& 4604.05 & 0.0330 &  0.0037 & CIV 1548    & 1.974 \nl
& 4611.76 & 0.0206 &  0.0032 & CIV 1550    & 1.974 \nl
& 4641.08 & 0.0146 &  0.0022 & FeII 2382   & 0.948 \nl
& 4642.48 & 0.0400 &  0.0028 & FeII 2382   & 0.948 \nl
76  \nl
& 4665.04 & 0.0986 &  0.0052 & CIV 1548    & 2.014 \nl
& 4666.98 & 0.5255 &  0.0064 & CIV 1548    & 2.014 \nl
& 4672.69 & 0.0517 &  0.0039 & CIV 1550    & 2.014 \nl
& 4674.83 & 0.3407 &  0.0057 & CIV 1550    & 2.014 \nl
& 4697.38 & B\tablenotemark{a}  & & FeII 1608   & 1.920 \nl
& 4705.00 & 0.1657 &  0.0066 & FeII 1611   & 1.920 \nl
75  \nl
& 4728.23 & 0.8145 &  0.0064 & FeII 2344   & 1.017 \nl
& 4731.12 & 0.0114 &  0.0022 & FeII 2344   & 1.018 \nl
& 4762.81 & 0.4712 &  0.0049 & C IV 1548   & 2.076 \nl
& 4770.56 & 0.2718 &  0.0041 & C IV 1550   & 2.076 \nl
74  \nl
& 4789.24 & 0.4111 &  0.0066 & FeII 2374   & 1.017 \nl
& 4805.90 & 1.1945 &  0.0052 & FeII 2382   & 1.017 \nl
& 4808.90 & 0.0340 &  0.0024 & FeII 2382   & 1.018 \nl
& 4840.87 & 0.0764 &  0.0032 & CIV 1548    & 2.128 \nl
& 4843.10 & 0.1480 &  0.0048 & CIV 1548    & 2.128 \nl
& 4843.84 & 0.0265 &  0.0039 & CIV 1548 ?  & 2.129 \nl
& 4844.75 & 0.1160 &  0.0053 & CIV 1548 ?  & 2.129 \nl
73  \nl
& 4848.88 & 0.0373 &  0.0035 & CIV 1550    & 2.128 \nl
& 4851.20 & 0.0775 &  0.0039 & CIV 1550    & 2.128 \nl
& 4852.74 & 0.0473 &  0.0049 & CIV 1550    & 2.129 \nl
& 4879.45 & 3.1863 &  0.0057 & AlII 1670   & 1.920 \nl
& 4888.16 & 0.0332 &  0.0031 &           &       \nl
& 4899.00 & 1.5708 &  0.0052 & MgII 2796   & 0.752 \nl
& 4911.72 & 1.4135 &  0.0055 & MgII 2803   & 0.752 \nl
72  \nl
& 4912.26 & 0.7060 &  0.0056 & MgII 2803   & 0.752 \nl
& 4947.94 & 0.0749 &  0.0024 & FeII 1608   & 2.076 \nl
& 4969.58 & 0.1531 &  0.0032 & CIV 1548    & 2.210 \nl
& 4973.91 & 0.0141 &  0.0024 & NiII 1703   & 1.920 \nl
& 4977.79 & 0.0957 &  0.0032 & CIV 1550    & 2.210 \nl
71  \nl
& 4992.82 & 0.3562 &  0.0068 & NiII 1709   & 1.920 \nl
& 4998.16 & 0.4040 &  0.0050 & Mg I 2852   & 0.752 \nl
70  \nl
& 5066.05 & 0.0417 &  0.0030 & FeII 2600   & 0.948 \nl
& 5079.82 & 0.1466 &  0.0029 & CIV 1548    & 2.281 \nl
& 5086.17 & 0.5020 &  0.0064 & NiII 1741   & 1.920 \nl
& 5088.39 & 0.1195 &  0.0034 & CIV 1550    & 2.281 \nl
& 5097.09 & 0.0114 &  0.0020 &           &       \nl
& 5116.46 & 0.3559 &  0.0071 & NiII 1751   & 1.920 \nl
69  \nl
& 5139.71 & 0.1553 &  0.0035 & AlII 1670   & 2.076 \nl
& 5141.45 & 0.1865 &  0.0068 & CIV 1548    & 2.321 \nl
& 5150.00 & 0.0679 &  0.0063 & CIV 1550    & 2.321 \nl
& 5197.93 & 0.0231 &  0.0040 & MnII 2576   & 1.017 \nl
68  \nl
& 5217.15 & 0.7297 &  0.0059 & FeII 2586   & 1.017 \nl
& 5233.60 & 0.0226 &  0.0029 & MnII 2594   & 1.017 \nl
& 5244.40 & 1.2150 &  0.0046 & FeII 2600   & 1.017 \nl
& 5247.70 & 0.0219 &  0.0020 & FeII 2600   & 1.018 \nl
& 5251.41 & 0.0148 &  0.0025 &           &       \nl
& 5257.58 & 0.0188 &  0.0029 & MnII 2606   & 1.017 \nl
& 5270.17 & 0.0177 &  0.0028 &           &       \nl
67  \nl
& 5280.20 & 0.9075 &  0.0093 & SiII 1808   & 1.920 \nl
& 5333.72 & 0.1860 &  0.0040 & CIV 1548    & 2.445 \nl
& 5342.63 & 0.0982 &  0.0042 & CIV 1550    & 2.445 \nl
66  \nl
& 5385    & 0.0210 &  0.0038 & TiII 3073   & 0.752 \nl
& 5394.39 & 0.0119 &  0.0021 &           &       \nl
& 5410.96 & 0.0150 &  0.0019 &           &       \nl
& 5411.26 & 0.0134 &  0.0016 &           &       \nl
& 5416.90 & 2.0526 &  0.0073 & AlIII 1854  & 1.920 \nl
65  \nl
& 5441.40 & 0.4677 &  0.0055 & AlIII 1862  & 1.920 \nl
& 5446.74 & 0.1397 &  0.0034 & MgII 2796   & 0.948 \nl
& 5448.36 & 0.3411 &  0.0028 & MgII 2796   & 0.948 \nl
& 5460.73 & 0.0932 &  0.0025 & MgII 2803   & 0.948 \nl
& 5462.30 & 0.2674 &  0.0026 & MgII 2803   & 0.948 \nl
& 5473.29 & 0.0125 &  0.0017 &           &       \nl
& 5483.81 & 0.0602 &  0.0026 & CIV 1548    & 2.542 \nl
& 5491.80 & B\tablenotemark{a}  & & CIV 1550    & 2.542 \nl
& 5492.87 & 0.0352 &  0.0024 & CIV 1550    & 2.542 \nl
64  \nl
& 5556.92 & 0.0131 &  0.0016 & MgI 2852    & 0.948 \nl
& 5575.17 & 0.0209 &  0.0024 & TiII 1910a  & 1.920 \nl
& 5578.96 & 0.0174 &  0.0024 & TiII 1910a  & 1.920 \nl
& 5580.05 & 0.0286 &  0.0026 & TiII 1910b  & 1.920 \nl
& 5581.14 & 0.0231 &  0.0027 & TiII 1910b  & 1.920 \nl
63  \nl
& 5640.23 & 2.0372 &  0.0055 & MgII 2796   & 1.017 \nl
& 5643.64 & 0.0907 &  0.0025 & MgII 2796   & 1.018 \nl
& 5654.54 & 1.7992 &  0.0046 & MgII 2803   & 1.017 \nl
& 5658.10 & 0.0528 &  0.0021 & MgII 2803   & 1.018 \nl
& 5660.88 & 0.0144 &  0.0017 & TiII 3230   & 0.752 \nl
& 5680    & 0.0590 &  0.0046 & TiII 3242   & 0.752 \nl
62  \nl
& 5749.35 & 0.0122 &  0.0020 &           &       \nl
& 5754.59 & 0.3604 &  0.0049 & MgI 2852    & 1.017 \nl
61  \nl
60  \nl
& 5897.44 & S\tablenotemark{b}  & & Sky Abs     &       \nl
& 5901.23 & S\tablenotemark{b}  & & Sky Abs     &       \nl
& 5902.81 & S\tablenotemark{b}  & & Sky Abs     &       \nl
& 5910.25 & S\tablenotemark{b}  & & Sky Abs     &       \nl
& 5917.00 & 0.3952 &  0.0069 & ZnII 2026   & 1.920 \nl
& 5919.57 & S\tablenotemark{b}  & & Sky Abs     &       \nl
& 5920.27 & S\tablenotemark{b}  & & Sky Abs     &       \nl
& 5923.73 & S\tablenotemark{b}  & & Sky Abs     &       \nl
& 5925.02 & S\tablenotemark{b}  & & Sky Abs     &       \nl
& 5930    & 0.0963 &  0.0045 & TiII 3384   & 0.752 \nl
& 5933.25 & S\tablenotemark{b}  & & Sky Abs     &       \nl
& 5942.36 & S\tablenotemark{b}  & & Sky Abs     &       \nl
& 5942.90 & S\tablenotemark{b}  & & Sky Abs     &       \nl
& 5943.77 & S\tablenotemark{b}  & & Sky Abs     &       \nl
& 5950.38 & S\tablenotemark{b}  & & Sky Abs     &       \nl
& 5972.59 & S\tablenotemark{b}  & & Sky Abs     &       \nl
59  \nl
& 6005.24 & 0.4274 &  0.0072 & CrII 2056   & 1.920 \nl
& 6023.17 & B\tablenotemark{a}  & & CrII 2062   & 1.920 \nl
& 6034.29 & 0.2268 &  0.0059 & CrII 2066   & 1.920 \nl
58  \nl
57  \nl
& 6276 $-$  & S\tablenotemark{b}  & & Sky Abs     &       \nl
56  \nl
55  \nl
& 6439.81 & 0.2705 &  0.0047 &           &       \nl
& 6476.38 & S\tablenotemark{b}  & & Sky Abs     &       \nl
& 6477.06 & S\tablenotemark{b}  & & Sky Abs     &       \nl
& 6481.37 & S\tablenotemark{b}  & & Sky Abs     &       \nl
& 6484.54 & S\tablenotemark{b}  & & Sky Abs     &       \nl
& 6494.24 & S\tablenotemark{b}  & & Sky Abs     &       \nl
& 6497.18 & S\tablenotemark{b}  & & Sky Abs     &       \nl
& 6517.85 & S\tablenotemark{b}  & & Sky Abs     &       \nl
\tablenotetext{a}{ Line blending}
\tablenotetext{b}{ Night-sky absorption or emission features}
\enddata
\normalsize
\end{deluxetable}

\begin{deluxetable}{lcc}
\tablewidth{0pc}
\tablenum{4} 
\tablecaption{METAL TRANSITIONS}
\tablehead{
\colhead{Transition} & \colhead{$\lambda_{\rm rest}$ ($\AA$)} &
\colhead{$f$}}

\tiny
\startdata
SiIV 1393 & 1393.755 & 0.528 \nl
SnII 1400 & 1400.400 & 0.71 \nl
SiIV 1402 & 1402.770 & 0.262 \nl
GaII 1414 & 1414.402 & 1.8 \nl
SiII 1526 & 1526.707 & 0.2303 \nl
CIV 1548 & 1548.195 & 0.1908 \nl
CIV 1550 & 1550.770 & 0.09522 \nl
GeII 1602 & 1602.4863 & 0.135 \nl
FeII 1608 & 1608.4449 & 0.05545 \nl
FeII 1611 & 1611.2004 & 0.001020 \nl
AlII 1670 & 1670.7874 & 1.88 \nl
PbII 1682 & 1682.15 & 0.156 \nl
NiII 1741 & 1741.549 & 0.1035 \nl
NiII 1751 & 1751.910 & 0.0638 \nl
SiII 1808 & 1808.0126 & 0.00218 \nl
AlIII 1854 & 1854.716 & 0.539 \nl
AlIII 1862 & 1862.790 & 0.268 \nl
TiII 1910a & 1910.6 & 0.0975 \nl
TiII 1910b & 1910.97 & 0.0706 \nl
ZnII 2026 & 2026.136 & 0.515 \nl
CrII 2056 & 2056.254 & 0.1403 \nl
CrII 2062 & 2062.234 & 0.1049 \nl
ZnII 2062 & 2062.664 & 0.2529 \nl
FeII 2344 &  2344.214 & 0.1097 \nl
FeII 2374 & 2374.4612 & 0.02818 \nl
FeII 2382 & 2382.765 &  0.3006 \nl
FeII 2600 & 2600.1729 & 0.2239 \nl
MgII 2796 & 2796.352 & 0.6123 \nl
MgII 2803 & 2803.531 & 0.3054 \nl
TiII 3073 &  3073.877  & 0.1091 \nl
TiII 3230 & 3230.131 & 0.05861 \nl
TiII 3242 & 3242.929 & 0.1832 \nl
TiII 3384 & 3384.740 & 0.3401  \nl
\enddata
\normalsize
\end{deluxetable}

\begin{deluxetable}{ccccclcc}
\tablewidth{0pc}
\tablenum{5a} 
\tablecaption{FIT FOR $z$=1.920 -- LOW IONS}
\tablehead{
\colhead{Comp} & \colhead{$z$} &
\colhead{$\sigma_z$} & \colhead{$b$} &
\colhead{$\sigma_b$} & \colhead{ION} &
\colhead{log $N$} & \colhead{$\sigma_{\log N}$} \\
& & \colhead{$\sci{-5}$} & \colhead{(\kms)} &
\colhead{(\kms)} & & \colhead{($\cm2$)} & \colhead{($\cm2$)}}

\tiny
\startdata
 1 & 1.919912 &  4.4 & 10.28 &  2.71 & Si$^+$     & 14.81 &  0.23 \nl
 & & & & & Fe$^+$     & 14.35 &  0.27 \nl
 & & & & & Ni$^+$     & 12.77 &  0.25 \nl
 & & & & & Cr$^+$     & 12.70 &  0.23 \nl
 & & & & & Zn$^+$     & 11.89 &  0.24 \nl
 2 & 1.919991 &  0.2 &  5.04 &  0.68 & Si$^+$     & 15.00 &  0.14 \nl
 & & & & & Fe$^+$     & 14.65 &  0.12 \nl
 & & & & & Ni$^+$     & 13.14 &  0.10 \nl
 & & & & & Cr$^+$     & 12.84 &  0.15 \nl
 & & & & & Zn$^+$     & 12.09 &  0.14 \nl
 & & & & & Mg$^0$     & 12.00 &  0.17 \nl
 3 & 1.920169 &  0.8 &  9.66 &  1.62 & Si$^+$     & 14.72 &  0.07 \nl
 & & & & & Fe$^+$     & 14.30 &  0.12 \nl
 & & & & & Ni$^+$     & 12.77 &  0.07 \nl
 & & & & & Cr$^+$     & 12.37 &  0.10 \nl
 & & & & & Zn$^+$     & 11.51 &  0.12 \nl
 4 & 1.920407 &  1.7 & 11.76 &  2.00 & Si$^+$     & 14.97 &  0.11 \nl
 & & & & & Fe$^+$     & 14.68 &  0.11 \nl
 & & & & & Ni$^+$     & 13.02 &  0.12 \nl
 & & & & & Cr$^+$     & 12.83 &  0.12 \nl
 & & & & & Zn$^+$     & 12.16 &  0.11 \nl
 5 & 1.920566 &  0.5 &  3.93 &  1.33 & Si$^+$     & 14.71 &  0.14 \nl
 & & & & & Fe$^+$     & 14.23 &  0.20 \nl
 & & & & & Ni$^+$     & 12.72 &  0.18 \nl
 & & & & & Cr$^+$     & 12.53 &  0.19 \nl
 & & & & & Zn$^+$     & 11.90 &  0.13 \nl
 6 & 1.920660 &  0.5 &  1.51 &  1.93 & Si$^+$     & 14.70 &  0.34 \nl
 & & & & & Fe$^+$     & 14.37 &  0.12 \nl
 & & & & & Ni$^+$     & 12.50 &  0.13 \nl
 & & & & & Cr$^+$     & 12.31 &  0.15 \nl
 & & & & & Zn$^+$     & 11.64 &  0.11 \nl
 7 & 1.920706 &  3.8 & 17.91 &  4.63 & Si$^+$     & 15.16 &  0.16 \nl
 & & & & & Fe$^+$     & 14.77 &  0.19 \nl
 & & & & & Ni$^+$     & 13.27 &  0.16 \nl
 & & & & & Cr$^+$     & 12.66 &  0.12 \nl
 & & & & & Zn$^+$     & 12.29 &  0.16 \nl
 8 & 1.921033 &  3.0 & 19.72 &  2.85 & Si$^+$     & 14.85 &  0.10 \nl
 & & & & & Fe$^+$     & 14.61 &  0.11 \nl
 & & & & & Ni$^+$     & 13.06 &  0.09 \nl
 & & & & & Cr$^+$     & 13.10 &  0.15 \nl
 & & & & & Zn$^+$     & 12.06 &  0.09 \nl
 9 & 1.921373 &  0.8 & 10.77 &  2.97 & Si$^+$     & 14.11 &  0.12 \nl
 & & & & & Fe$^+$     & 13.94 &  0.27 \nl
 & & & & & Ni$^+$     & 12.11 &  0.15 \nl
 & & & & & Cr$^+$     & 11.64 &  0.31 \nl
 & & & & & Zn$^+$     & 10.83 &  0.48 \nl
\enddata
\normalsize
\end{deluxetable}

\begin{table*}
\dummytable\tablenum{3}\label{abs}
\end{table*}

\begin{table*}
\dummytable\tablenum{4}\label{osc}
\end{table*}

\begin{table}
\dummytable\tablenum{5a}\label{1920L}
\end{table}

\begin{table*}
\begin{center}
\begin{tabular}{ccccclcc}
Comp & $z$ & $\sigma_z$ & b & $\sigma_b$ & Ion & log $N$ &
$\sigma_{\rm{log} {\it N}}$\cr
&
& ($\sci{-5}$)
& (\kms)
& (\kms)
&
& ($\cm2$)
& ($\cm2$) \cr
\tableline
 1 & 1.919534 &  0.6 &  8.56 &  0.76 & Al$^{++}$  & 12.22 &  0.03 \cr
 2 & 1.919733 &  0.7 &  9.27 &  1.66 & Al$^{++}$  & 12.17 &  0.06 \cr
 3 & 1.919993 &  0.3 & 10.02 &  0.34 & Al$^{++}$  & 13.35 &  0.01 \cr
 4 & 1.920245 &  0.3 & 12.69 &  1.04 & Al$^{++}$  & 13.28 &  0.03 \cr
 5 & 1.920397 &  0.3 &  3.63 &  0.91 & Al$^{++}$  & 12.47 &  0.09 \cr
 6 & 1.920597 &  0.2 & 13.70 &  0.45 & Al$^{++}$  & 13.21 &  0.01 \cr
 7 & 1.920898 &  1.1 &  5.57 &  2.18 & Al$^{++}$  & 12.42 &  0.19 \cr
 8 & 1.921050 &  1.2 & 13.05 &  1.63 & Al$^{++}$  & 12.78 &  0.06 \cr
 9 & 1.920804 &  1.3 &  3.62 &  1.70 & Al$^{++}$  & 12.19 &  0.19 \cr
10 & 1.921285 &  0.6 &  5.25 &  1.38 & Al$^{++}$  & 11.98 &  0.08 \cr
11 & 1.921431 &  0.4 &  6.28 &  0.53 & Al$^{++}$  & 12.30 &  0.02 \cr
\cr
 1 & 1.920053 &  7.4 & 44.25 &  3.19 & C$^{+3}$   & 14.47 &  0.11 \cr
 2 & 1.920327 &  0.6 & 16.77 &  2.17 & C$^{+3}$   & 16.80 &  0.68 \cr
 3 & 1.920906 &  2.2 & 27.52 &  2.39 & C$^{+3}$   & 14.12 &  0.05 \cr
 4 & 1.921361 &  0.2 &  8.97 &  0.41 & C$^{+3}$   & 13.96 &  0.02 \cr
 5 & 1.921688 &  1.4 & 11.86 &  2.66 & C$^{+3}$   & 13.09 &  0.17 \cr
 6 & 1.921743 &  2.0 & 30.92 &  1.57 & C$^{+3}$   & 13.82 &  0.05 \cr
\end{tabular}
\end{center}

\tablenum{5b}
\caption{FIT FOR $z$=1.920 -- HIGH IONS} \label{1920H}
\end{table*}

\clearpage

\begin{table*}
\begin{center}
\begin{tabular}{ccccclcc}
Comp & $z$ & $\sigma_z$ & b & $\sigma_b$ & Ion & log $N$ &
$\sigma_{\rm{log} {\it N}}$\cr
&
& ($\sci{-5}$)
& (\kms)
& (\kms)
&
& ($\cm2$)
& ($\cm2$) \cr
\tableline
 1 & 2.076214 &  0.1 &  5.75 &  0.23 & Fe$^{+}$   & 13.34 &  0.02 \cr
 & & & & & Al$^{+}$   & 12.19 &  0.01 \cr
\cr
 1 & 2.076155 &  0.8 & 11.30 &  0.66 & C$^{+3}$   & 13.45 &  0.03 \cr
 2 & 2.076301 &  1.4 &  5.66 &  2.39 & C$^{+3}$   & 12.84 &  0.32 \cr
 3 & 2.076422 &  2.0 &  9.50 &  1.63 & C$^{+3}$   & 13.22 &  0.10 \cr
 4 & 2.076707 &  1.5 &  8.72 &  2.71 & C$^{+3}$   & 12.21 &  0.09 \cr
\cr
 1 & 2.076132 &  0.3 &  4.77 &  0.49 & Si$^{+3}$  & 12.55 &  0.02 \cr
 2 & 2.076393 &  0.6 & 11.03 &  1.00 & Si$^{+3}$  & 12.54 &  0.03 \cr
\end{tabular}
\end{center}

\tablenum{6}
\caption{FIT FOR $z$=2.076 -- LOW AND HIGH IONS} \label{2076F}
\end{table*}

\begin{table*}
\begin{center}
\begin{tabular}{ccccclcc}
Comp & $z$ & $\sigma_z$ & b & $\sigma_b$ & Ion & log $N$ &
$\sigma_{\rm{log} {\it N}}$\cr
&
& ($\sci{-5}$)
& (\kms)
& (\kms)
&
& ($\cm2$)
& ($\cm2$) \cr
\tableline
 1 & 0.947779 &  0.1 &  2.23 &  0.17 & Mg$^{+}$   & 12.64 &  0.04 \cr
 & & & & & Fe$^{+}$   & 11.85 &  0.05 \cr
 2 & 0.947861 &  0.5 &  4.15 &  1.72 & Mg$^{+}$   & 11.47 &  0.07 \cr
 3 & 0.948103 &  0.6 &  1.51 &  3.86 & Mg$^{+}$   & 11.14 &  0.09 \cr
 4 & 0.948336 &  0.1 &  3.12 &  0.27 & Mg$^{+}$   & 13.09 &  0.06 \cr
 & & & & & Fe$^{+}$   & 12.00 &  0.06 \cr
 5 & 0.948389 &  0.4 &  6.11 &  0.42 & Mg$^{+}$   & 12.69 &  0.04 \cr
 & & & & & Fe$^{+}$   & 11.92 &  0.07 \cr
\end{tabular}
\end{center}

\tablenum{7}
\caption{FIT FOR $z$=0.948} \label{0948F}
\end{table*}

\clearpage

\begin{table*}
\begin{center}
\begin{tabular}{ccccclcc}
Comp & $z$ & $\sigma_z$ & b & $\sigma_b$ & Ion & log $N$ &
$\sigma_{\rm{log} {\it N}}$\cr
&
& ($\sci{-5}$)
& (\kms)
& (\kms)
&
& ($\cm2$)
& ($\cm2$) \cr
\tableline
 1 & 1.016661 &  0.3 &  3.63 &  1.01 & Fe$^{+}$   & 12.80 &  0.04 \cr
 & & & & & Mg$^{0}$  & 10.82 &  0.07 \cr
 2 & 1.016783 &  3.9 &  6.71 &  5.55 & Fe$^{+}$   & 12.94 &  0.41 \cr
 & & & & & Mg$^{0}$  & 10.78 &  0.46 \cr
 3 & 1.016839 &  0.7 &  3.04 &  1.80 & Fe$^{+}$   & 13.08 &  0.26 \cr
 & & & & & Mg$^{0}$  & 11.04 &  0.24 \cr
 4 & 1.016935 &  0.3 &  4.59 &  1.24 & Fe$^{+}$   & 12.91 &  0.05 \cr
 & & & & & Mg$^{0}$  & 11.04 &  0.06 \cr
 5 & 1.017026 &  0.6 &  4.02 &  2.21 & Fe$^{+}$   & 12.48 &  0.11 \cr
 & & & & & Mg$^{0}$  & 10.83 &  0.10 \cr
 6 & 1.017156 &  0.1 &  6.10 &  0.26 & Fe$^{+}$   & 14.00 &  0.01 \cr
 & & & & & Mg$^{0}$  & 11.93 &  0.01 \cr
 7 & 1.017256 &  0.2 &  5.63 &  0.29 & Fe$^{+}$   & 13.74 &  0.02 \cr
 & & & & & Mg$^{0}$  & 11.70 &  0.02 \cr
\end{tabular}
\end{center}

\tablenum{8}
\caption{FIT FOR $z$=1.107} \label{1017F}
\end{table*}

\begin{table*}
\begin{center}
\begin{tabular}{ccccclcc}
Comp & $z$ & $\sigma_z$ & b & $\sigma_b$ & Ion & log $N$ &
$\sigma_{\rm{log} {\it N}}$\cr
&
& ($\sci{-5}$)
& (\kms)
& (\kms)
&
& ($\cm2$)
& ($\cm2$) \cr
\tableline
 1 & 2.013144 &  0.9 & 19.21 &  1.24 & C$^{3+}$   & 12.95 &  0.02 \cr
 2 & 2.014375 &  0.1 &  7.99 &  0.25 & C$^{3+}$   & 13.71 &  0.01 \cr
 & & & 5.23 & 0.25 & Si$^{3+}$  & 12.95 &  0.02 \cr
 3 & 2.014408 &  2.3 & 38.72 &  3.30 & C$^{3+}$   & 13.32 &  0.04 \cr
 & & & 25.32 & 3.30 & Si$^{3+}$  & 12.38 &  0.06 \cr
 4 & 2.014544 &  0.7 &  5.90 &  1.01 & C$^{3+}$   & 12.80 &  0.06 \cr
 5 & 2.014900 &  2.6 & 13.64 &  3.95 & C$^{3+}$   & 12.40 &  0.17 \cr
\end{tabular}
\end{center}

\tablenum{9}
\caption{FIT FOR $z$=2.014} \label{201F}
\end{table*}

\clearpage

\begin{table*}
\begin{center}
\begin{tabular}{lcccc}
Transition
& Apparent
& VPFIT
& EqW
& Adopted \cr
\tableline
FeII 1611  & $ 15.461 \pm  0.020 $ & $ 15.454 \pm  0.058 $ & & $15.458 \pm
0.058$ \cr
NiII 1709  & $ 13.902 \pm  0.010 $ & $ 13.880 \pm  0.054 $ & & \cr
NiII 1741  & $ 13.878 \pm  0.006 $ & \cr
NiII 1751  & $ 13.912 \pm  0.010 $ & \cr
SiII 1808  & $ 15.794 \pm  0.005 $ & $ 15.807 \pm  0.059 $ & & $15.801 \pm
0.059$ \cr
CrII 2056  & $ 13.628 \pm  0.009 $ & $ 13.645 \pm  0.062 $ & & $13.753 \pm
0.062$ \cr
CrII 2066  & $ 13.681 \pm  0.013 $ & \cr
ZnII 2026  & $ 12.935 \pm  0.009 $ & $ 12.912 \pm  0.057 $ & & $12.924 \pm
0.057$ \cr
\cr
CuII 1358  & $12.992 \pm 0.063$ & & $ 12.929 \pm 0.068$ & $12.962 \pm 
0.065$ \cr
GeII 1602  & $< 12.430$ & & $< 12.431$ & $< 12.430$ \cr
TiII 1910  & & & $12.807 \pm 0.038$ & $12.807 \pm 0.038$ \cr
\cr
C IV 1548  & & $16.804 \pm 0.675$ & & $16.804 \pm 0.675$ \cr
\cr
AlIII 1854 & $13.874 \pm 0.003$ & $13.887 \pm 0.013$ & & $13.881 \pm
0.013$ \cr
\end{tabular}
\end{center}

\tablenum{10}
\caption{IONIC COLUMN DENSITIES FOR $z$ = 1.920} \label{1920I}

\tablecomments{Values reported in logarithmic space have deceptively 
small errors.  For
instance, the value for $\N{Cu}$ is not a $5 \sigma$ detection.} 

\end{table*}

\clearpage

\begin{table*}
\begin{center}
\begin{tabular}{lcccc}
Transition
& Apparent
& VPFIT
& EqW
& Adopted \cr
\tableline
FeII 1608  & $ 13.296 \pm  0.018 $ & $ 13.342 \pm  0.016 $ & &
$13.320 \pm 0.016$ \cr
AlII 1670  & $ 12.174 \pm  0.012 $ & $ 12.194 \pm  0.014 $ & &
$12.184 \pm 0.014$ \cr
SiII 1304  & $ 13.756 \pm  0.037 $ & & & $13.756 \pm 0.037$ \cr
C II 1334  & $ 14.151 \pm  0.069 $ & & & $> 14.151$ \cr
NiII 1751  & $< 12.289$ & & $< 12.305$ & $< 12.297$ \cr
CrII 2056  & $ 12.12 \pm  0.12 $ & & $12.10 \pm 0.13$ &
$12.11 \pm 0.13$ \cr
ZnII 2026  & $ < 11.330 $ & & $< 11.342$ & $< 11.336$ \cr
\cr
Si IV 1402  & $ 12.833 \pm  0.019 $ & $ 12.848 \pm  0.017 $ \cr
C IV 1548  & $ 13.713 \pm  0.005 $ & $ 13.727 \pm  0.055 $ \cr
\end{tabular}
\end{center}

\tablenum{11}
\caption{IONIC COLUMN DENSITIES FOR $z$ = 2.076} \label{2076I}

\end{table*}

\clearpage

\begin{table*}
\begin{center}
\begin{tabular}{lcc}
Transition
& Apparent
& Adopted \cr
\tableline
FeII 2374  & $ 14.611 \pm  0.030 $ & $ 14.611 \pm  0.030 $ \cr
FeII 2260  & $ 14.743 \pm  0.084 $  \cr
MgI 2852   & $ 12.364 \pm  0.006 $ & $ 12.364 \pm  0.006 $ \cr
MnII 2606  & $ 12.420 \pm  0.053 $ & $ 12.532 \pm  0.053 $ \cr
MnII 2594  & $ 12.386 \pm  0.051 $ \cr
MnII 2576  & $ 12.641 \pm  0.018 $ \cr
TiII 3073  & $ 12.243 \pm  0.067 $ & $12.300 \pm 0.030 $ \cr
TiII 3242  & $ 12.319 \pm  0.034 $ \cr
TiII 3384  & $ 12.228 \pm  0.021 $ \cr
\end{tabular}
\end{center}

\tablenum{12}
\caption{IONIC COLUMN DENSITIES FOR $z$ = 0.752} \label{0752I}

\end{table*}

\begin{table*}
\begin{center}
\begin{tabular}{lccc}
Transition
& Apparent
& VPFIT
& Adopted \cr
\tableline
FeII 2382  & $ 12.318 \pm  0.050 $ & $ 12.401 \pm  0.034 $ &
$12.401 \pm 0.034$ \cr
FeII 2586  & $ 11.936 \pm  0.391 $ &  \cr
FeII 2600  & $ 12.452 \pm  0.046 $ &  \cr
MgII 2796  & $ 13.069 \pm  0.008 $ & $ 13.342 \pm  0.038 $ & 
$13.342 \pm 0.038$ \cr
MgII 2803  & $ 13.146 \pm  0.006 $ & \cr
MgI 2852   & $ 10.81 \pm  0.15 $ & & $10.81 \pm 0.15$ \cr
\end{tabular}
\end{center}

\tablenum{13}
\caption{IONIC COLUMN DENSITIES FOR $z$ = 0.948} \label{0948I}

\end{table*}

\begin{table*}
\begin{center}
\begin{tabular}{lccc}
Transition
& Apparent
& VPFIT
& Adopted \cr
\tableline
FeII 2344  & $ 14.168 \pm  0.086 $ &  \cr
FeII 2374  & $ 14.242 \pm  0.008 $ & $ 14.285 \pm  0.026 $ & 
$14.285 \pm 0.026$ \cr
FeII 2586  & $ 14.212 \pm  0.007 $ & $ 14.285 \pm  0.026 $ \cr
MgI 2852   & $ 12.236 \pm  0.007 $ & $ 12.247 \pm  0.024 $ & $12.247 
\pm 0.024$ \cr
\end{tabular}
\end{center}

\tablenum{14}
\caption{IONIC COLUMN DENSITIES FOR $z$ = 1.017} \label{1017I}

\end{table*}

\clearpage

\begin{table*}
\begin{center}
\begin{tabular}{lccc}
Metal
& log$_{10} N$(X) (cm$^{-2}$)
& [X/H] \hfil \cr
\tableline
Fe & $15.458 \pm 0.058$ & $-0.705 \pm 0.097$ \cr
Ni & $13.891 \pm 0.054$ & $-1.012 \pm 0.095$ \cr
Si & $15.801 \pm 0.059$ & $-0.402 \pm 0.098$ \cr
Cr & $13.753 \pm 0.062$ & $-0.580 \pm 0.100$ \cr
Ti & $12.807 \pm 0.038$ & $-0.776 \pm 0.081$ \cr
Zn & $12.924 \pm 0.057$ & $-0.379 \pm 0.097$ \cr
\cr
Cu & $12.962 \pm 0.065$ & $0.039 \pm 0.102$ \cr
Ge & $< 12.430$ & $< 0.147$ \cr
\end{tabular}
\end{center}

\tablenum{15}
\caption{ABUNDANCES FOR $z$ = 1.920} \label{1920A}
\end{table*}

\begin{table*}
\begin{center}
\begin{tabular}{lccc}
Metal
& log$_{10} N$(X) (cm$^{-2}$)
& [X/H] \hfil \cr
\tableline
Fe & $13.320 \pm 0.016$ & $-2.621 \pm 0.071$ \cr
Al & $12.184 \pm 0.014$ & $-2.727 \pm 0.070$ \cr
Si & $13.756 \pm 0.037$ & $-2.225 \pm 0.075$ \cr
Ni & $< 12.297$ & $< -2.384$ \cr
Cr & $12.11 \pm 0.13$ & $-2.00 \pm 0.15$ \cr
Zn & $< 11.336$ & $< -1.745$ \cr
C  & $> 14.151$ & $> -2.840$ \cr
\end{tabular}
\end{center}

\tablenum{16}
\caption{ABUNDANCES FOR $z$ = 2.076} \label{2076A}
\end{table*}

\clearpage

\end{document}